\documentclass[useAMS,usenatbib]{mn2e}
\usepackage{graphicx}
\usepackage{latexsym,epsfig}
\usepackage{longtable,rotating}
\usepackage[utf8]{inputenc}

\usepackage{lscape}
\usepackage{array,arydshln}

\title[The Hercules stream]{Chemical compositions and kinematics of the Hercules stream}
\author[Ramya, Reddy, Lambert and Musthafa]{P. Ramya,$^{1}$\thanks{E-mail:
ramyap09@gmail.com} Bacham E. Reddy$^{1}$, David L. Lambert$^{2}$ and M. M. Musthafa$^{3}$ \\
$^{1}$Indian Institute of Astrophysics, Bengaluru, India-560034\\
$^{2}$The W.J. McDonald Observatory \& Department of Astronomy,University of Texas at Austin, Austin, TX 78712, USA \\
$^{3}$University of Calicut, Malappuram, India-673635 }

\begin{document}

\date{Accepted ..... Received .... ; in original form ....}

\pagerange{\pageref{firstpage}--\pageref{lastpage}} \pubyear{...}

\maketitle

\label{firstpage}

\begin{abstract}
An abundance analysis is reported of 58 K giants identified 
by Famaey et al. (2005) as highly probable members of the Hercules 
stream selected from stars north of the celestial equator in 
the {\it Hipparcos} catalogue. 
The giants have compositions spanning the interval [Fe/H] from $-0.17$ to 
$+0.42$ with a mean value of $+0.15$ and relative elemental abundances [El/Fe] 
representative of the Galactic thin disc. Selection effects may have 
biassed the selection from the {\it Hipparcos} catalogue against the 
selection of metal-poor stars. Our reconsideration of 
the recent extensive survey of FG 
dwarfs which included metal-poor stars  (Bensby et al. 2014) provides a 
[Fe/H] distribution for the Hercules stream 
which is  similar to that from the 58 giants.  It appears that the stream is dominated by
metal-rich stars from the thin disc. 
Suggestions in the literature that the stream   includes  metal-poor stars from the thick disc
are discussed.
 
\end{abstract}


\section{Introduction}

The local Milky Way disc is resolvable into two principal large-scale components -- the thin and the thick disc -- 
with distinct differences in
chemical composition,  kinematics and stellar ages.   
Within the disc there  are open clusters comprised of  stars of a common age and composition moving together
around the centre of the Galaxy. These clusters are not permanent features but disperse due to their internal velocity
dispersion and interaction with the gravitational field of the Galaxy: see, for example, Feltzing \& Holmberg (2000). 
Dissolution of clusters feeds the population of field stars within the disc.   
Other aggregations of stars within
the disc have been identified through their common  kinematics. 
Various labels -- moving groups, stellar streams and superclusters -- have been applied to collections of
such kinematically-associated stars.  Modern investigations of these entities began with Eggen (1958a, b and c) who
 supposed that a moving
group was a  dissolving open cluster (see Eggen 1996 and references therein).
Today, proposed origins of stellar streams include not only this early idea of a dissolving cluster 
but the possibility  that some may be remnants of an accreted
dwarf satellite galaxy (Navarro, Helmi \& Freeman 2004) and that others   
 are collections of field stars with common kinematics arising from 
dynamical perturbations within the Galaxy including the
effects of the spiral arms  (e.g., Quillen \& Minchev 2005)  and/or the central bar (Fux 2001). 
For fuller discussion of possible origins of the stellar
streams see Antoja et al. (2010) and Sellwood  (2014).

A kinematic search for stellar streams and determination of stream members follows Eggen's pioneering
ideas and  involves 
quantitative scrutiny  of the Galactic motions  $(U,V,W)$, where $U$, $V$ and $W$ 
are the radial, tangential and vertical components of velocity.   
In assigning membership of a stream, the three components may be considered directly or in simple combinations.
With the publication of the {\it Hipparcos} catalogue,  
several authors compiled $(U,V,W)$ components for various stellar  samples. In Figure 1,
we show the $U,V$-plane for local K and M giants as presented by Famaey et al. (2005, their Figure 9).   
Their sample of about 6000 giants, all north of the celestial equator and believed to be
single stars,  was
sorted by a maximum likelihood method into several classes including the three most striking moving groups  --
Hyades-Pleiades, Sirius  and Hercules  streams -- as well as
young giants, high-velocity (thick disc and halo) giants 
and a group of about 4000 background (mostly thin disc) giants.  The fractions of  the 6030 stars
assigned to the streams are 7.0\%. 5.3\% and 7.9\% for the Hyades-Pleiades, Sirius and Hercules streams, respectively. 
This paper is devoted to the chemical compositions of those stars assigned to the Hercules stream.

 Eggen (1958a  and c), based on a set of 700 high-velocity stars 
identified a number of  stars with kinematic motions similar to $\zeta$ Herculis, a bright star 
in the Hercules constellation. Hence, 
 the group is called the Hercules  stream.  
A major characteristic of the Hercules
stream in the solar neighbourhood is its negative $U$ velocity, i.e., it is moving outward from the Galactic centre.
In a 2008 paper (Famaey et al. 2008) based on  a wavelet approach, the Hercules stream 
-- the green elongated structure in Figure 1 -- was resolved into two clumps corresponding to
$U$ from $-65$ to $-49$ km s$^{-1}$ and $U$ from $-40$ to $-30$ km s$^{-1}$, 
both with $V$ from $-55$ to $-47$ km s$^{-1}$. 
In an  analysis of  Famaey et al.'s sample of giants and of the Geneva-Copenhagen survey of
14,000 dwarfs (Nordstr\"{o}m et al. 2004), Zhao, Zhao \& Chen (2009) 
identified 22 moving groups  including the Hercules stream. 
(Throughout this paper unless otherwise stated, velocities $U,V,W$ are heliocentric values.)

 Our principal goal is to present abundance
analyses of  giants from Famaey et al.'s list of very probable members of the Hercules stream.
In the following  section (Section 2), we discuss the sample and our spectroscopic observations. 
Section 3 covers  the abundance analysis, the
elemental abundances,  stellar ages and kinematics. 
Section 4 discusses our abundances in the context of the compositions of the
Galactic thin and thick discs as well as previous determinations of the compositions of 
Hercules stream members, notably that by Bensby et al.
(2007, 2014) for dwarf members of the stream.
The concluding Section 5 summarizes our principal results concerning
the nature of the Hercules stream.

\section{Sample and Observations}

Our sample of giants was chosen from  Famaey et al. (2005). 
By applying a maximum-likelihood method to the kinematic
data of more than 6000 K and M giants (excluding binaries) they grouped their sample into a 
number of substructures  (Figure 1) and  identified 529 stars as 
belonging to the Hercules stream  with 204 
given a high probability  ($>$ 70\%) of membership in the stream.  
It was this latter subset that forms the basis of our sample.
To avoid very cool stars with complex spectra 
and to restrict the sample to stars with  atmospheric parameters similar 
to the bright K giant star, Arcturus, we chose stars with 
the colour cut off  $V-I$ $<$ 1.2. Further, this  sample of 140 stars was culled to be accessible from
the Harlan J. Smith 2.7~m Telescope at the W.J. McDonald Observatory at the time of our observing runs.

Spectra of  58 giants in the Hercules stream were obtained
using the Robert G. Tull coud\'{e} spectrograph (Tull et al. 1995) at the
Harlan J. Smith Telescope. Observed spectra cover  the
wavelength range: 3800-10000 \AA. However, beyond $\sim$ 5800 \AA, coverage is incomplete owing to 
gaps between the recorded portions of the spectral orders which
progressively increases towards redder wavelengths.
At a resolving power of $\sim$ 60,000, 
the observed spectra have a S/N of 100 or more over much of the spectrum.
Spectral images were reduced to one dimensional 
flux versus wavelength using the software 
Image Reduction and Analysis Facility (IRAF).\footnote{IRAF is distributed by the National 
Optical Astronomy Observatory, which is operated by the Association of Universities for 
Research in Astronomy (AURA) under cooperative agreement with the National Science Foundation.}
The pre-processing tasks such as the bias and flat corrections, wavelength calibration using ThAr 
lamp spectrum, scattering correction, continuum fitting and Doppler correction were 
performed on the data.

\section{Analysis}

In this section, we  describe photometric and spectroscopic estimations of atmospheric 
parameters,  the derivation of elemental abundances, ages  and  
kinematical properties of the observed giants from the Hercules stream.

\subsection{Stellar Atmospheric Parameters}

Atmospheric parameters effective temperature $T_{\rm eff}$, surface gravity  log $g$, 
microturbulence  $\xi_{t}$ and metallicity [M/H]  were obtained 
for each star using spectral line analysis combined with stellar models. 
For deriving atmospheric parameters 
we used convective stellar model atmosphere with no convective overshoot (Kurucz 1998) and the  2009 version 
of the LTE line analysis  code MOOG (Sneden 1973). 
Line data were taken from the compilations of Reddy et al. (2003) 
and Ram\'{\i}rez \& Allende Prieto (2011). 

 An iterative method of spectroscopic estimation of the stellar atmospheric parameters was followed. The first step 
in this process is to choose an element with numerous lines in the spectra  and having a good range in 
line strengths, lower excitation Potential (LEP) and also having adequate numbers
 of both neutral and ionized lines. 
 The element Fe satisfying these conditions was chosen and Table 1 gives the list 
of Fe\,{\sc i}  lines (LEP in the range of 0-5 eV) and Fe\,{\sc ii} lines used in the analysis.  The effective temperature 
was set by the requirement that the 
Fe abundance provided by Fe\,{\sc i} lines  be independent of their LEPs. 
This was achieved by forcing the slope of Fe\,{\sc i} abundances versus LEPs plot to be  zero. 
$\xi_{t}$ is derived based on the fact that it affects lines of different strengths differently. 
The weak lines are unaffected by microturbulence 
while the strong lines are highly affected. $\xi_{t}$ is fixed such that Fe\,{\sc i} abundances are independent of their 
line strengths by demanding the slope of the abundances versus log($W_{\lambda}$/$\lambda$) plot  be zero.
As the number of Fe\,{\sc ii} lines were 
 fewer than for  Fe\,{\sc i} lines, less weightage was given to them while deriving $T_{\rm eff}$ and $\xi_{t}$.  
Also, while deriving $T_{\rm eff}$, care was taken to 
minimize the effect of $\xi_{t}$ by choosing, initially very weak lines with a sufficient range in LEP. Later, 
$\xi_{t}$ was fixed by adding moderately strong Fe\,{\sc i} lines. The surface gravity log $g$ was obtained by requiring 
that for the given $T_{\rm eff}$ and $\xi_{t}$, Fe \,{\sc i} and Fe\,{\sc ii} lines give the same abundance. This mean 
abundance of Fe is considered to be the metallicity [M/H]. Derived parameters based on the spectroscopic analysis 
are given in Table~2.

The uncertainties of the estimated values of stellar atmospheric parameters are found by checking 
systematically, how  the Fe abundances used for their determination respond to the changes in their values. 
To find the  uncertainty in $T_{\rm eff}$, the temperature was varied in steps of $\pm$ 25 K by keeping all other 
parameters - log $g$, $\xi_{t}$ and [M/H] - constant;  for a change of $\pm$ 50 K, the slope of the Fe \,{\sc i} abundance 
versus LEP plot changed 
appreciably. The uncertainty in  log $g$ was estimated by varying the log $g$ value 
keeping $T_{\rm eff}$, $\xi_{t}$ and [M/H] constant,  until  the Fe\,{\sc ii} abundance  differed noticeably  from the Fe\,{\sc i} 
abundance value; this led to a 
measured uncertainty of $\pm$50 K and $\pm$0.2 cm s$^{-2}$ in 
$T_{\rm eff}$ and log $g$, respectively. A similar analysis by changing the $\xi_{t}$, keeping the other parameters constant  gave an 
uncertainty 
of $\pm$0.2 km s $^{-1}$ in $\xi_{t}$, corresponding to a visually detectable slope to the Fe\,{\sc i} abundance versus 
log($W_{\lambda}$/$\lambda$) plot. 
These individual uncertainties translate to an effective error of $\pm$0.1 dex in metallicity. It should be noted that 
the errors in the stellar parameters are not uncorrelated, as the same lines or subsets of the same set of lines are used to 
derive all of them. This exercise was done for a representative 
star HIP 8926 and the estimated uncertainties in model parameters are reprentative of all the stars in our sample.

Derived spectroscopic values of $T_{\rm eff}$ and log $g$ were checked against those  derived using photometry.
$T_{\rm eff}$ values were derived using empirical calibrations of optical and infrared photometry. 
Two colours $J-K$ and $V-K$ and the colour-temperature relations 
given in Alonso, Arribas \& Mart\'{\i}nez-Roger (1999)
 were used. 
Magnitudes of $J$ and $K$ were taken from 
2MASS sky survey (Cutri et al. 2003). 
The 2MASS magnitudes were converted to the standard TCS system using 
the calibration given in Ram\'{\i}rez \& Mel\'{e}ndez (2005). 
The $V$ magnitude in $V-K$ was taken from Tycho2 catalogue. 
Values of $V_{T}$ magnitudes in the Tycho2 catalogue, were converted to Johnson broad band magnitude $V$ using 
the relation provided in the  Tycho2 catalogue. The $K_{s}$ magnitudes in 2MASS were converted to $K$ magnitudes 
of the Johnson system using the relation in Bessell (2005). 
Magnitudes were corrected for interstellar extinction using the relations 
given in Ram\'{\i}rez \& Mel\'{e}ndez (2005).  
We assumed [Fe/H] = 0 for relations that contain the
$V-K$ colour. Derived
$T_{\rm eff}$ values from $J-K$ and $V-K$ colour are given in Table~2, and  
are compared in Figure~2. An error of 40 K and 125 K respectively, are associated with the photometric temperatures 
($T_{\rm eff}$)$_{V-K}$ and ($T_{\rm eff}$)$_{J-K}$ estimated using $V-K$ and $J-K$ colour respectively. 
Spectroscopic and  photometric temperatures agree quite well but problems exist for the coolest stars.  For the warmer stars,  the
photometric temperatures are about 200 K cooler than the spectroscopic values.

The surface gravities (log $g$) were obtained through web interface (PARAM code\footnote {http://stev.oapd.inaf.it/cgi-bin/param\_1.3}) 
for the Bayesian estimation of stellar parameters (see da Silva et al. 2006 for details), using the 
PARSEC isochrones (Bressan et al. 2012). 
The lognormal initial mass function from Chabrier (2001) and a constant star formation rate were assumed as Bayesian priors. 
As input parameters, ($T_{\rm eff}$)$_{spec}$, [Fe/H], parallax (van Leeuwen 2007) 
and dereddened $V$ magnitude were used. Note that, we derived extinction values E(B-V) using the extinction 
maps of Schlegel et al. (1998) and the recipes provided in Bonifacio et al. (2000). Derived values are tabulated in Table~2 and 
are compared in Figure~2. 
As shown in the Figure, log $g$ values from photometry are systematically lower by about 0.2 dex compared to values 
from spectroscopy, which is consistent with the results given in da Silva et al. (2006). Given the uncertainties in the parallaxes, 
photometry and the spectroscopic determination, 
the agreement between the two methods is satisfactory.  In this study, we used the atmospheric parameters
derived from the spectroscopy.

\subsection{The Elemental Abundances}

In this section,  we present abundances of 15 elements  for the  58 giants of the Hercules stream and compare results with
those from two independent and smaller surveys of Hercules giants also extracted from the Famaey et al. sample.

The measured lines and
adopted atomic data are given in Table~3. Abundances were
derived using the Kurucz stellar model atmospheres and the LTE line analysis 
 code $MOOG$.
Our reference solar abundances
were derived using the solar spectrum taken from Hinkle et al. (2000). The measured equivalent width of each line 
and the estimated abundances from these lines in Sun are given in Table 3. The standard spectroscopic procedure
as outlined in section 3.1   
was used to obtain solar atmospheric parameters : $T_{\rm eff}$ = 5835 K, log $g$ = 4.55 cm s$^{-2}$, 
$\xi_{t}$ = 1.25 km s$^{-1}$ for the metallicity [M/H] = 0. 
The derived solar abundances along with the literature  values
(Asplund et al. 2009)  are given 
in Table~4. Our and the literature results agree within the estimated uncertainties.
Derived abundances for the 58 Hercules giants are given in Table~5 and Table 6 in the form [Fe\,{\sc i}/H], [Fe\,{\sc ii}/H] and 
then [El/Fe]  with two
entries for [Ti/Fe] : one for [Ti\,{\sc i}/Fe] and another for [Ti\,{\sc ii}/Fe]. The  Ti entries provide a check on
the spectroscopic $\log g$ which is based on equality of the Fe abundances from Fe\,{\sc i} and
Fe\,{\sc ii} lines.  The mean difference [Ti\,{\sc i}/Fe] $-$ [Ti\,{\sc ii}/Fe] is a satisfactory  $-0.02$ with a hint of 
more positive values or a larger
scatter at  the coolest effective temperatures.

Hyperfine corrections were applied for the elements Mn, Co, Sc and V using the wavelengths and relative strengths of 
hyperfine components taken from the  Kurucz database. Ba lines are  affected by both hyperfine and isotopic splitting. 
Ba has seven isotopes $^{130}$Ba, $^{132}$Ba, $^{134}$Ba, $^{135}$Ba, $^{136}$Ba, $^{137}$Ba and $^{138}$Ba 
contributing 0.106 \%, 0.101 \%, 2.417 \%, 6.592 \%, 7.854 \%, 11.23 \% and 71.70 \% of the total solar system 
Ba abundance respectively (Anders \& Grevesse 1989). We adopted the wavelength and relative strength of 
Ba  components from McWilliam (1998). The Zn abundance was based initially on  the
 transitions at  4810.54 \AA\ and  6362.35 \AA. However, measurement of $W_{\lambda}$ of 
the transition at  4810.54 \AA\ was found to be unreliable for some stars as 
one wing  overlaps with  Fe\,{\sc ii} and/or Cr\,{\sc i} lines (Hinkle et al.  2000). 
Thus, we list Zn abundance only when both the lines were measured.

The error in the mean abundance can be statistically represented as $\overline{\sigma}$ = $\sigma$/$\sqrt{N}$, where 
$\sigma$ represents the standard deviation or the line to line spread in the abundances and N is the number of lines used for the 
abundance estimation (see Table 5 and 6 for the values). But, the systematic uncertainties associated with the 
estimated stellar parameters and with the atomic data and measured equivalent width of the spectral lines, 
get propagated to the estimated abundance value, and exceed the $\overline{\sigma}$ in most of the cases. 
The resultant uncertainty in the abundances due to any parameter is gauged by measuring the amount by which the mean abundances 
vary responding to a change in the respective parameter by an amount equal to the uncertainty in the parameter.  
Following the method, the resultant uncertainties in the abundances, $\sigma$($T_{\rm eff}$), $\sigma$(log $g$),  
$\sigma$($\xi_{t}$) and $\sigma$([M/H]) due to the uncertainties in the atmospheric parameters - $\Delta$($T_{\rm eff}$) = $\pm$ 50 K 
in temperature, $\Delta$(log $g$) = $\pm$ 0.20 cm s$^{-2}$ in surface gravity, $\Delta$($\xi_{t}$) = $\pm$ 0.20 km s$^{-1}$ 
in microturbulent velocity and $\Delta$([M/H]) = $\pm$ 0.10 dex in metallicity respectively, were estimated. The atomic data we are 
mainly concerned with  are  the log $gf$ values, which are carefully chosen from literature. Moreover, 
errors in $gf$-values get cancelled out in the differential abundance analysis with respect to Sun, as the same lines with 
same atomic data 
are used for both. The uncertainty in the measured $W_{\lambda}$ is estimated using the recipe given in Cayrel (1988) and 
it turned out to be in the range from 1m{\AA} to about 2.5m{\AA} for our data. 
On average, the uncertainty in the measured $W_{\lambda}$ is taken to be $\Delta(W_{\lambda}$) = $\pm$ 2m{\AA}.
As there are N lines used for the analysis, the quantity $\overline{\sigma}$($W_{\lambda}$) = $\sigma$(W$_{\lambda}$)/$\sqrt{N}$ 
represents the 
error in abundances due to $\Delta(W_{\lambda}$). Although these uncertainties are not uncorrelated, assuming 
them to be so, they are added in quadrature to get the total systematic error in the abundances.

The above procedure was applied to a typical member (HIP 8926) of our Hercules stream sample and the results are 
given in Table 7. Assuming that the uncertainties in the parameters are unrelated and independent, the net 
uncertainty in the abundance ratio was calculated as a quadratic sum which is given 
in the last column in Table~7 as $\sigma_{model}$. Abundance ratios of singly ionised 
species Ti\,{\sc ii}, Fe\,{\sc ii} and Ba\,{\sc ii} were found to have large uncertainties due to their higher sensitivity to 
uncertainties in $\log g$ and $\xi_{t}$.

Two recent independent studies have  analyzed Hercules giants from the Famaey et al. sample using high-resolution
spectra of a similar quality to ours,  similar  analytical techniques but independent selections of lines.  These provide an external 
check on our
results.  Pakhomov,  Antipova \& Boyarchuk  (2011) obtained spectra of
17 giants of which 11 are in common with our sample. Their selected giants have a greater than 80\% probability of belonging to the
Hercules stream.  Liu et al. (2012)  observed 10 giants of which two are in common with
us.  Pakhomov et al. and Liu et al. have no stars in common.

For the Pakhomov et al. sample, differences in abundance  between 
the two 
studies ($\Delta$[X/H] = current value - Pakhomov value)  for common stars 
 are given in Table~8. 
Agreement  is, 
in general, satisfactory except  for one or two of the 11 common stars. 
For example, for HIP 107502  the
abundance values are systematically higher in our study. This  is likely due  to our 
higher value of 
$T_{\rm eff}$ by 290~K and $\log g$ by 1.1 dex compared to 
Pakhomov et al. (2011). Also, one notices 
significantly higher values 
of Mn and Co for most of the stars in our study. It is not clear whether the difference 
is  due to different  $\log gf$-values or to differing treatment 
of HFS in the two studies. 
Pakhomov et al. (2011)  considered HFS splitting for Co but not for Mn.  We include HFS for both these elements.
Additionally, there is incomplete overlap in the lines of Mn\,{\sc i}  and Co\,{\sc i} chosen for the two abundance analyses.  
There  are large differences for 
Ba\,{\sc ii}.  A major contributor here is that the $\log gf$-values adopted by Pakhomov et al. are 0.2 dex larger than our 
values for the two
lines we use to obtain the Ba abundance.  Pakhomov et al.'s sample have a probability of greater than 80\% of belonging to the
Hercules stream; recall that our sample was chosen to have a probability of 70\% or greater.

For the two stars in common with Liu et al., the mean difference for [X/Fe] from 11 common elements is 0.03 for one 
(HIP 51047) and 0.02 for the other star (HIP 37441) with 10 of 11 or 9 of the 11 [El/Fe]  falling within $\pm0.10$ dex.  The [Fe/H]
values  differ by 0.11 dex for HIP 51047 and 0.09 dex for HIP 37441 in the sense that our values are larger.  In part the [Fe/H]
differences are likely attributable to our hotter $T_{\rm eff}$ and stronger surface gravity.  
Liu et al. adopted a looser criterion for membership than either ourselves or Pakhomov et al.; 
the probabilties for membership are as low as 45\%
and 46\% for two stars and only four of their ten stars meet our criterion of a probability of 70\% or greater.

The histogram for [Fe/H] for our 58 giants is shown in Figure  3.  The sample  spans the [Fe/H] range from $-$0.17 to $+$0.43 with 
a peak at [Fe/H] = $+$0.15.  
A characteristic of the stream is its spread in $U$ (Figure 1) but the composition of the stream's
giants does not appear to depend on $U$; stars with $U < -50$ km s$^{-1}$ and those with $U > 50$ km s$^{-1}$
share the same [Fe/H] as the total sample and the two samples  are not distinguishable by their [El/Fe].
The  [Fe/H] distribution is
quite similar to that from Pakhomov et al.'s  sample: their mean [Fe/H] from 17 stars is $+0.11$.   Liu 
et al.'s quartet of stars with a 70\% or greater probability of stream membership have 
[Fe/H] of $-0.36$, $-0.26$, $-0.25$ and $0.0$ for a mean
of $-0.22$   which, if we adjust by the $0.1$  difference between their and 
our results from two common stars (see above), gives a mean
of $-0.12$  a value not too severely different from our mean value.  In summary,  these two independent  analyses of Hercules
giants provide compositions generally similar to results from our larger sample.

Two characteristics of Figure 3 surely deserve comment.  First,  metal-poor stars, say [Fe/H] $\leq -0.5$ are absent. Second,  
observed stars
span a [Fe/H] range much greater than the measurement uncertainty and, therefore,  indicate that the Hercules 
stream is not a dissolving open cluster.   
A contributing factor to the absence of metal-poor stars may be a selection effect embedded in the Famaey et al. sample (2005).
Their principal selection criteria were that  a star must have a spectral type of K or M in the {\it Hipparcos} input catalogue  and then
an absolute magnitude limit was imposed  to separate the K (and M) giants from dwarfs.  
Given that evolutionary tracks for giants move to the blue and earlier
spectral types with decreasing metallicity, metal-poor giants will not have been selected by Famaey et al.   The cut-off as
the spectral type moves to G-type for giants  probably occurs at slightly less than [Fe/H] of $-0.5$; Arcturus 
with [Fe/H] $\simeq -0.5$ has a spectral
type of  K2IIIp in the {\it Hipparcos} catalogue.  Secondly,  the local density of metal-poor field stars is low and, 
if  a stream's population  even
approximately mirrors that of the
field,  very large or targeted samples are surely required 
in order to locate 
metal-poor stars in the Hercules stream.  
The density of metal-poor giants  in a distance-limited sample such as the {\it Hipparcos}
catalogue is likely lower than for metal-poor dwarfs because of the shorter lifetime of a giant.  In short,  absence 
of metal-poor giants  in the
Hercules stream as selected by Famaey et al. (2005)  likely arises from selection effects and may not be an intrinsic property of the stream.  
Such selection effects may be mitigated and even eliminated by surveys selected by  $U,V,W$ velocities.

Confirmation of  a metallicity distribution centred at about the solar metallicity is provided by two analyses which draw a 
large fraction of their stellar
sample from the Geneva-Copenhagen survey of FGK dwarf stars  (Nordstr\"{o}m et al. 2004; Holmberg, Nordstr\"{o}m \& Andersen  et al. 2007).  
Antoja et al. (2008) identify 
a sample of  Hercules stream  members among  the more 10000 FGK dwarfs with  $(U,V)$ values very similar to those isolated by
Famaey et al. (2005).  With Str\"{o}mgren metallicities from the Geneva-Copenhagen catalogue,  Antoja et al.  (their Table 4 and Figure 15)
give the mean [Fe/H] as $-0.15\pm0.27$.  Bobylev, Bajkova \& Myll\"{a}ri (2010)  with a sample of more than 15000 stars identity  
almost 200 stars with the
Hercules stream  distributed among two clumps  coincident with Famaey et al.'s identification of the stream in Figure 1.  
Mean metallicities, again from
Str\"{o}mgren  metallicities,  are [Fe/H] $= -0.09\pm0.17$ for the clump at $(U,V) = (-33, -51)$ and $-0.16\pm0.27$ for 
the clump at $(U,V) =  (-77,-49)$ and
these values not surprisingly are consistent with those from Antoja et al.   Even these two samples provide too few members 
of the Hercules stream to
contain metal-poor stars.

\subsection {Ages}

Ages of individual stars were estimated using the PARAM code and PARSEC isochrones (Bressan et al. 2012) with Bayesian 
priors for the Initial mass function (the lognormal function from Chabrier 2001) and the Star Formation Rate (constant). 
Essential background to the PARAM code is provided by da Silva et al. (2006) who remark that derived quantities such as the 
mean stellar age and its standard error are derived from a probability distribution function for the logarithm of the age. 
Spectroscopic effective temperature and [Fe/H] were used
as input along with van Leeuwen's revised {\it Hipparcos} parallax and the reddening corrected $V$ magnitude.  Ages and error  
estimates are given in Table 2. A histogram of stellar ages is shown in Figure~4.  Except for two outliers 
(HIP 7710 at 6.1$\pm$2.8 Gyr and HIP 48417 at 7.1$\pm$2.6 Gyr), the ages are less than 4 Gyr with a peak around 1.5 Gyr.  
The width of the histogram exceeds typical errors and, thus, confirms that the stream is not a dissolved open cluster.  
These ages are typical of local thin disc stars (Bensby et al. 2005,  2014; Reddy et al. 2003) and not of the local thick disc 
for which far greater ages have been derived, as, for example, the 10.7 Gyr by Reddy et al. and 9.7 Gyr by Bensby et al.

Age estimates for common stars were compared with Pakhomov et al. (2011) who estimated ages from  isochrones (Girardi et al. 2000) 
and a star's location in the $\log  L/L_{\odot}$ versus $T_{\rm eff}$ plane.  Agreement with our estimates is very good except for 
two stars: HIP 107502 where our age is much shorter and HIP 108012 where our age is much longer than estimates given by 
Pakhomov et al. For the former case, a possible explanation is that our effective temperature is nearly 300 K warmer which reduces 
the age and also brings the abundances into agreement.

Ages have also been estimated for FG dwarfs identified with the Hercules stream.  With age estimates for stars in the 
Geneva-Copenhagen survey derived from Str\"{o}mgren photometry  and a Bayesian estimation technique, Antoja et al. (2008)  
found an age distribution peaking at 2.2 Gyr with a full-width and half power of about 1.2 Gyr with a low density tail extending 
to at least 8 Gyr (see their Figure 17). This age profile is similar to that illustrated in Figure 4.  Adopting age estimates from 
Holmberg  et al. (2007),  Bobylev et al. (2010)  quote ages of around 5 Gyr with an uncertainty of about 3 Gyr for both cores of 
the Hercules stream. These estimates appear incompatible with those from Antoja et al. and our values as summarized in Figure 4.

\subsection{Kinematics and Stream Membership}

In light of van Leeuwen's (2007) revisions to the {\it Hipparcos} astrometry, we decided to recalculate the motions $(U,V,W)$ of our giants.
Radial velocities were estimated from our spectra by cross-correlation with a template spectrum. The typical error 
in radial velocity is $\pm$1 km s$^{-1}$. The measured radial velocities in this study agree well with those 
given by Famaey et al. (2005);  the mean difference (ours - them)  is  
$+$0.47 $\pm$1.25 km s$^{-1}$. Space velocities $(U,V,W)$ were derived  
using the recipe given in Johnson \& Soderblom (1987) combined with new radial velocities and the 
updated  {\it Hipparcos} astrometry. 
The mean differences between the values derived in this study and those
 given Famaey et al. (2005) 
are small: $-$1$\pm$8, 0.3$\pm$7 and 0.2$\pm$3 km s$^{-1}$, respectively for $U$, $V$ and $W$.
These small differences  have no more than a minor effect on the selection of Hercules stream members from
Famaey et al.'s catalogue.

Crucial to the identification of members of the Hercules stream  is the procedure for isolating stream members
from  other stars in the $(U,V)$ plane and/or other diagrams involving the $(U,V,W)$ velocities or quantities
derived from these velocity components.   Our sample of giants drawn from Famaey et al. (2005)  adopts their
probability estimates for stream membership based on the mean stream velocities of
$(U,V,W) = (-42,-52,-8)$ and their $(\sigma_U,\sigma_v,\sigma_W)$ - see their Table 2 for a full specification of
the parameter set including the fraction of their total sample assigned to the Hercules stream and to the other
components (see above): 7.9$\pm0.9$ \% of the 6030 stars are assigned to the Hercules stream.  Our selection criterion
was that stars must have a 70\% or greater probability of belonging to the Hercules stream, as estimated by Famaey et al. (2005).
It is assumed that Famaey et al.'s (2008)  subsequent resolution of the $U$ distribution of the Hercules stream into two
clumps ($U$ from $-49$ to $-65$ km s$^{-1}$ and $-30$ to $-40$ km s$^{-1}$ with $V$ from $-47$ to $-55$ km s$^{-1}$)
does not greatly affect the probability of stream membership.  Probability calculations performed by Famaey et al. (2005, 2008)
are similar  but not exactly equivalent to the  calculations for Galactic entities  described by Bensby et al. (2003),
Reddy et al. (2006) and Bensby et al. (2014) based on Gaussian distribution functions for the different
entities.  Use of  Reddy et al.'s (2006) recipe with Famaey et al.'s (2005) parameters for their six  populations  overestimates the
probability that a star belongs to the Hercules stream but by only a few per cent.

\section{The Hercules stream and the Galactic disc}

In essentially all respects, the composition of the Hercules giants is identical to that of thin disc giants in the solar  neighborhood. 
Luck \& Heiter's (2007) abundance analyses of a large sample of local  GK giants is our reference sample. The vast majority of their
giants belong to the thin Galactic disc;  discussion of possible differences between thin and thick disc stars is given below.
In terms of the [Fe/H] distribution, Luck \& Heiter's sample has a mean value [Fe/H] = 0.0  and a sharp cutoff at [Fe/H] $\simeq +0.3$
and a low [Fe/H] tail extending to $-0.6$. Our sample of giants has a higher mean [Fe/H] ($= +0.15$) and a higher fraction of high [Fe/H]
stars   than this local field sample and, as noted above,  a distinct paucity of metal-poor stars which likely results in part from 
the selection effects
inherent in the Famaey et al. (2005) sample.  (A sample of local FGK dwarfs analyzed by Luck \& Heiter (2006) has a more
pronounced high [Fe/H] tail and closely resembles our histogram for [Fe/H] $\ge 0.0$.)  With the possible exception of our higher
fraction of `very' metal-rich stars,  the [Fe/H] distribution for Hercules giants resembles that of local giants.  An excess of 
high metallicity stars among  the Hercules stream may imply origins in the inner Galaxy  where  metallicities are higher at about the
gradient of 0.1 dex kpc$^{-1}$ according to measured abundance gradients.  As noted above, metal-poor giants are probably 
underrepresented in these histograms.

The concordance between [Fe/H] distributions for Hercules and local field giants extends to element ratios [El/Fe].  In Table 9,
we give [El/Fe]  for the three [Fe/H] intervals: [Fe/H] $= -0.2$ to $0.0$, $0.0$ to $+0.2$ and $+0.2$ to $+0.4$ for our sample of Hercules
giants and Luck \& Heiter's (2007) field giants.  Inspection of Table 9 shows a close correspondence between [El/Fe] values in the
two samples across the full [Fe/H] interval.  This is further illustrated in Figures  5 and 6. Figure 5 shows [Na/Fe] versus [Fe/H] for 
the Hercules giants, Luck \& Heiter's (2007) giants and FG dwarfs from Bensby et al. (2014).
Figure 6 shows [Mg/Fe], [Si/Fe] and [Ni/Fe] versus [Fe/H]
for our sample of Hercules giants and local giants from Luck \& Heiter (2007). 
Slight differences in [El/Fe] may arise
from differences in analytical techniques  but the impression is that the [El/Fe] values for the Hercules stream are the same
as those for local field giants which represent the Galactic thin disc.

Several discussions of the Hercules stream have suggested
that the stream has a component from the Galactic thick disc.  Differences in composition between thin and thick discs are most
obvious at low [Fe/H], say [Fe/H] $\leq -0.3$, where, in particular, the abundances of $\alpha$-elements Mg, Si, Ca and Ti are enhanced in
the thick relative to the thin disc with [$\alpha$/Fe] $\simeq +0.3$. Differences between thin and thick discs decrease as [Fe/H] = 0
is approached.  The local thick disc is represented by stars with [Fe/H] as low as $-1.2$ and extends up to 0.0 (Bensby et al. 2007) and
possibly to  $+0.2$ (Bensby et al. 2014). The interval [Fe/H] $\simeq 0.0$ is populated also by a class of stars dubbed high-$\alpha$ 
metal-rich (HAMR)
stars introduced by Adibekyan  et al. (2011) who proposed that  HAMR stars formed a separate population in the local disc. Bensby et al.
(2014), on the other hand,  suggest  HAMR stars are simply the metal-rich extension of the thick disc.  For our present purpose,
the relation between the HAMR stars and the metal-rich thick disc is academic. What is key is that there are differences between the
composition of thin and of thick/HAMR stars  at the same [Fe/H]  but such differences  diminish as [Fe/H] = 0 is approached from low
[Fe/H].   This merger is shown in Figure 7 for [Ti/Fe]  where results for our Hercules giants are shown with results for `certain'  thin 
and thick
disc dwarfs from Bensby et al. (2014).  For [Fe/H] $\leq  0.0$, a thick (HAMR) -- thin disc separation is present  and the Hercules giants 
favour the
thin disc.  Note that high [Ti/Fe]  and low [Ti/Fe] loci are not populated purely by  kinematically-selected thin and thick disc stars, 
respectively;
this mixing suggests either the labels `thin' and `thick' are misapplied in some cases and/or the indices [Ti/Fe] and [$\alpha$/Fe]  are 
not a guaranteed
identifier of a population. Bensby et al.'s Figure 23  which shows  [Ti/Fe] versus [Fe/H]   with stars differentiated by their high or 
low [Ti/Fe] value for each
[Fe/H] value shows the HAMR stars with [Ti/Fe] excesses of about 0.15, 0.05 and 0.0 dex at [Fe/H] of $-0.2, 0.0$ and $+0.2$ dex, 
respectively, over
thin disc stars with [Ti/Fe] of close to 0.0.   By a very large majority, the Hercules giants in Figure 7 fall with the thin disc and 
not the HAMR
stars.   A cautionary note: the HAMR-thin disc abundances from Bensby et al. and Adibekyan et al. are derived from standard analyses
of   warm dwarfs but our results are for giants of cooler temperatures   and, hence, systematic differences in [El/Fe] might intrude 
into Figure 7.
Luck \& Heiter's 
(2006, 2007) standard analyses of such  dwarfs and giants show that differences are small  and, in addition, standard
analyses of dwarfs by Luck \& Heiter and by Bensby et al. are in fine agreement (Reddy, Giridhar \& Lambert
2015 -- their Table 16).  Across the [Fe/H] interval covered by the Hercules stream giants, their compositions indicate  that they 
closely resemble thin
disc stars rather than the thick disc/HAMR stars.

The presence  of thick disc  stars, especially metal-poor examples, within the Hercules stream  was
advocated by 
Bensby et al.'s (2007) who concluded that the Hercules stream
was not `a unique Galactic stellar population but rather is a mixture of thin and thick disc stars' and this claim
was reinforced by Bensby et al. (2014)  who tightened
their 2007 criteria for stream membership: the 2014 criteria for
stream membership were that the
kinematic probabilities of membership in the thin disc ($D$), thick disc ($TD$) and the
Hercules stream ($Her$) should satisfy the constraints  $Her/D >2$ and $Her/TD > 2$
but in the 2007 paper the criteria were that these
two ratios should be greater than one.
However, the tougher criteria for membership did not materially alter the range in composition of the
stars identified with the Hercules stream.
The 2014 sample of some 30 members  range in [Fe/H] from $-0.9$ to $+0.3$ with almost all of the stars
having [Fe/H]  $\leq -0.2$ and the $\alpha$-element enhancement
([$\alpha$/Fe] $\geq +0.1$) characteristic of the thick disc.
One might  suggest that the Bensby et al. (2014) sample of the Hercules stream  arises {\it in toto}
from the thick disc!

This sample's composition contrasts sharply with ours:
we have no stars more metal-poor than [Fe/H] $\simeq -0.2$ and their relative abundances [El/Fe] are typical of the thin disc.
The contrast is not traceable to the fact that one analysis refers to
dwarfs and the other to giants.
Rather, it arises from different $(U,V,W)$ definitions for the
Hercules stream.
 Bensby et al. (2007, 2014) centre the stream at $(U,V) = (-51,-62)$
which is about 10 km s$^{-1}$ more negative in $U$ and $V$ than
the Famaey et al. (2005 -- see the green swath in Figure 1) values and the consensus of  pre- and post-2005 
determinations.\footnote{Bensby et al. (2014) give  velocities relative to the local standard
of rest (LSR). Using their {adopted values of Sun's velocity components relative to the LSR}, we straightforwardly obtain the heliocentric
velocities for their adopted centres of the Hercules stream and velocities for individual stars.}
%Selection of stream members is necessarily tied to the adoption of the
%kinematical parameters for the populations adopted as representing the local stars.

Given that the stream is stretched out in $U$ but rather tightly confined in $V$,
a crucial quantity in selection of members is the choice of $V$.
A consequence of  Bensby et al.'s choice for the  $(U,V)$ centroid is that
their  proposed stream members  populate the $(U,V)$ plane about 10 km s$^{-1}$ in $V$
below the green swath identified with the Hercules stream in Figure 1.
Thus,  Bensby et al.  place the
stream deeper among the thick disc stars than Famaey et al.'s choice of $(U,V)$ and
this, we suspected,  explains the large contribution from the thick disc to the
Bensby et al. selection of stream members.

To test our suspicion, we interrogated Bensby et al.'s (2014) list of 714 stars using  Famaey et al.'s
parameters for their six contributors to the local stellar population (see above)
and identified 28 stars having a probability of greater than 80\% of belonging
to the stream. This sample has a narrower [Fe/H] range than  that selected by Bensby et al.
-- see Figure 3 where we find our selection from Bensby et al.'s 714 stars has a [Fe/H]
distribution fairly similar to ours for the Hercules giants. In addition, the relative
abundances [El/Fe] for the  revised sample of  Hercules dwarfs  are similar to those for our Hercules giants.
 Notably, the [Ti/Fe] for the members from Bensby et al.  are overwhelmingly the low values
typical of the thin disc.  
Just three stars have the higher [Ti/Fe] with also higher [Mg/Fe], [Si/Fe] and [Ca/Fe]  seen among thick
and HAMR stars. Each of the three is somewhat Fe-poor ([Fe/H] from $-0.18$ to $-0.35$). Three  thick disc stars out of a sample
of 28 is not unexpected given that  the selection criterion  was a probability of stream membership of 80\% or greater. 
(As noted above, our calculation of probabilities appears to
overestimate slightly the probability of stream membership with respect to the probabilities given by Famaey et al. (2005) 
from their maximum likelihood method. )
The age distribution of the 25 low [Ti/Fe] stars is consistent with a thin disc origin:  20 stars with
reliable ages (see Bensby et al. 2014 who  adopt the criterion that the upper and lower estimates of a star's age should not  
exceed 4 Gyr)  provide a
mean age of 5.1 Gyr with a distribution from 1.4 Gyr to 10.4 Gyr. Thus,  Bensby et al.'s (2014)  large sample of FG dwarfs 
covering a  range of metallicities and ages when sifted with appropriate $(U,V)$  confirms that the 
Hercules stream is predominantly a feature of the thin disc with a majority of members having a near-solar
metallicity.

Prior to Bensby et al.'s search for members of the Hercules stream,
Soubiran \& Girard (2005) assembled  data for more than 700 dwarfs and subgiants.
By a technique involving Gaussian distribution functions
(Bensby et al. 2003, 2014; Reddy et al. 2003) Soubiran \& Girard
computed a star's probability of belonging to the thin disc, the thick disc and the Hercules
stream.
Such probabilities rely on the characterization of the Gaussian distribution
functions and the relative strength of the three populations. Soubiran \& Girard's choice
of the Gaussian distribution for the Hercules stream was taken from
Famaey et al. (2005) and for the thin and thick disc from Soubiran et al. (2003).
%The relative populations
%were set at 72\%, 19\% and 9\% for the thin disc,
%the thick disc and the Hercules stream, respectively.
Nearly 600 stars were assigned such probabilities.
Just five stars ranging in [Fe/H] from $-0.90$ to $+0.06$ had a probability of greater
than 80\% of belonging to the Hercules stream and three of the five have $\alpha$-enhancements
typical of the thick disc..
If we apply our criterion of a probability of membership of 70\% or greater for the Hercules stream,
the mean [Fe/H] from 21 stars is $-0.26$ and the distribution is weighted to  metal-poor
stars more severely than shown by our giants and also by the revised selection from Bensby et al.
Of the 21 stars, the great majority belong to the thin disc according to their [$\alpha$/Fe]
values.
Mishenina et al.
(2013) who discuss a survey of 276 FGK dwarfs of which slightly more than half were included in
Soubiran \& Girard's sample reach similar conclusions about the Hercules stream and its mix of thin and thick disc stars.

\section{Concluding remarks}

Our abundance analysis of 58 of the giants assigned to the Hercules stream by Famaey et al. (2005) shows
general agreement with Pakhomov et al.'s (2011) independent analysis of 17 of Famaey et al.'s giants.   
In terms of chemical compositions, the principal results are that (i) the [Fe/H] of the 58 giants are
spread across the interval $-0.15$ to $+0.45$ with a mean of $+0.15$ and (ii) the relative abundances
[El/Fe] for elements such as the $\alpha$-elements  which  among the [Fe/H] $<0$ stars are a
discriminant between thin and thick disc stars, the results suggest the Hercules stream is composed primarily of thin disc stars.
 The ages of the Hercules giants
average about 1.4 Gyr with a real star-to-star spread. 
These ages are typical of local thin disc
stars (Bensby et al. 2005, 2014; Reddy et al. 2003) and not of the local thick disc for which far
greater ages have been derived, as, for example, the 10.7 Gyr by Reddy et al. and 9.7 Gyr by Bensby et al.
Not unexpectedly, the spreads in compositions and
in ages among stream members exclude the possibility of identifying the stream as a dissolved
open cluster.

What was unexpected  was the disagreement between our
results and  published claims
that the metallicities of  Hercules stream members extended down to [Fe/H] $\sim -1$ and contained
representatives of both the thin and thick discs. In the case of the most extensive available survey of
the compositions of F and G disc dwarfs (Bensby et al. 2014), we argue above that their
results for [Fe/H] and [El/Fe] are quite consistent with ours when the Hercules stream
members are selected from among the dwarfs using the same (actually, very similar) 
criteria as those used by Famaey et al. (2005); Bensby et al.'s (2007, 2014) selection of Hercules stream
members were based on inappropriate choices for the stream's central $(U,V)$ velocities.  
Soubiran \& Girard (2005) and Mishenina et al. (2013) also assembled samples of F and G dwarfs and
searched their samples for members of the Hercules stream. Their  conclusions about
the composition of stream members are not as consistent with ours as the adjusted sample from Bensby et al. (2014). In particular, 
they both  conclude
that the stream is composed of stars with a much larger [Fe/H] spread than we find;  there is little to no overlap between the histogram
provided by the Hercules giants and that from the high probability dwarf/subgiant members  isolated by Soubiran \& Girard (2005).  
There is
apparent agreement from composition (i.e., [$\alpha$/Fe]) and age on the point that the various samples are dominated by thin rather 
than thick disc stars.

\newpage

Differences in the [Fe/H] spread remain unresolved quantitatively; it is not obviously attributable to the choice of
kinematical parameters.\footnote{An idea that an explanation might be
found in the fact that the surveys sample different areas of the sky was short-lived. Famaey et al.'s
(2005) sample is drawn from stars in the {\it Hipparcos} catalogue north of the celestial equator. Soubiran \&
Girard's and Mishenina et al.'s samples, which do not support our abundance results,
are heavily weighted to the northern celestial hemisphere
while  Bensby et al.'s (2014) large sample, which supports our results, is greatly concentrated on the
southern celestial hemisphere.}  Factors contributing to differences between our sample of the stream and those offered by
Soubiran \& Girard and Mishenina et al.  may be the small number of high probability  members from the latter two surveys
and the biasses in our sample against the inclusion of metal-poor stars.  The fact  that our sample of stream giants and 
Bensby et al.'s sample of
stream dwarfs drawn from a selection of stars across a wide range in [Fe/H] are similar might suggest that the bias  inherent 
in the Famaey et al.
(2005) sample is small.   Perhaps most curious of all is the observation that the collection of stream members  extracted from 
Bensby et al.'s (2014)
large survey of FG stars is so different  in terms of their  [Fe/H] range from that provided by Sourbiran \& Girard's (2005) 
almost equally large survey of FG stars and with a quite similar distribution of stars by [Fe/H].
Such a contrast can not be attributed to systematic differences between abundance analyses nor to the technique from
separating stream members from thin and thick disc stars.

Together the compositions, the ages and the kinematics support the picture of the Hercules  stream as a flow of
stars driven outwards from the inner Galaxy by gravitational perturbations arising from the
rotating central bar (see, for example, Fux 2001). The outward flow is not uniform -- the $U$ velocities
span about 100 km s$^{-1}$ -- and unlikely to be steady over long periods because the gravitational perturbations from the Galactic 
bar and
spiral arms are time dependent. (Note that  a $U$ velocity of 50 km s$^{-1}$
maintained for 1 Gyr covers a distance of 51 kpc.) 
 Given that our [El/Fe] follow  closely
the thin  disc pattern, we suppose that the Hercules stream
is a thin and not a thick disc phenomenon, a conclusion confirmed by the estimates of stellar ages. Abundance analyses from the {\it APOGEE}
project show that the inner Galaxy along the Galactic plane is dominated by a thin disc-like
 (i.e., low [$\alpha$/Fe]) population
with the mean [Fe/H] increasing toward the Galactic centre (Hayden et al.  2015). Away from the Galactic
plane ($1 < |z| < 2$ kpc), a thick disc-like (i.e., high [$\alpha$/Fe]) population dominates 
the inner Galaxy with little
representation from the thin disc. Paucity of thick disc representatives in the Hercules stream
would suggest inefficient  mixing in height about the Galactic plane.    In contrast to the Hercules stream as a thin disc phenomenon, 
the two thick
disc streams -- Arcturus and AF06 -- analysed by Ramya et al. (2012)  have compositions representative of the thick disc without  
thin disc contamination. 
Observational insights  into the character of the Hercules stream and novel challenges to a fuller theoretical understanding will 
come as 
the  Hercules stream is mapped more extensively  in distance and azimuth as large surveys  are completed, e.g. Antoja et al. (2014)  
discuss the Hercules stream
as seen by the RAVE survey,  and much may be expected from scrutiny of {\it GAIA} results.

Further insights into  the formation and evolution of moving groups and streams may be expected from
detailed studies of additional streams in and beyond the solar neighbourhood. Zhao et al. (2009) identified 22 moving groups in the
solar neighbourhood. Of this total, few have been subjected to  comprehensive abundance analysis.
Pursuit of additional moving groups near and far is to be encouraged for as
Sellwood (2014) remarks `It is likely that the different streams have different origins 
and a combination of these [theoretical] ideas would be needed to explain all 
the features.'

\section{Acknowledgments}

We are most grateful to the referee for a thorough and constructive report on
our initial submission.
DLL thanks the Robert A. Welch Foundation of Houston, Texas for support through grant F-634.  We thank Alain Jorissen for a
helpful exchange of emails concerning Famaey et al.'s selection of Hercules giants and Tamara  Mishenina  for a clarification about 
the results from
the survey reported by Mishenina et al. (2013).
\\

%\begin{thebibliography}

{\bf REFERENCES } \\

Adibekyan V. Z., Santos N. C., Sousa S. G., Israelian G., 2011, A\&A, 535, L11 \\
Alonso A., Arribas S., Mart{\'{\i}}nez-Roger C., 1999, A\&AS, 140, 261 \\
Anders E., Grevesse N., 1989. Geochim. et Cosmochim. Acta, 53, 197 \\
Antoja T., Figueras, F., Fern\'{a}ndez, D., Torra, J., 2008, A\&A, 490, 135 \\
Antoja T., Figueras F., Torra J., Valenzuela O., Pichardo B., 2010, Lecture Notes and Essays in Astrophysics, 4, 13 \\
Antoja, T., Helmi, A., Dehnen, W. et al., 2014, A\&A, 563, A60 \\
Asplund M., Grevesse N., Sauval A. J., Scott P., 2009, ARA\&A, 47, 481 \\
Bensby T., Feltzing S., Lundstr\"{o}m I., 2003, A\&A, 410, 527 \\
Bensby T., Feltzing S., Lundstr\"{o}m I., Ilyin I., 2005, A\&A, 433, 185 \\
Bensby T., Oey M. S., Feltzing S., Gustafsson B., 2007, ApJ, 655, L89 \\
Bensby T., Feltzing S., Oey M. S., 2014, A\&A, 562, A71 \\
Bessell M. S., 2005, ARA\&A, 43, 293 \\
Bobylev, V.V., Bajkova, A.T., Myll\"{a}ri, A.A., 2010, Astr. Letters, 36, 27 \\
Bonifacio P., Monai S., Beers T. C., 2000, AJ, 120, 2065 \\
Bressan A., Marigo P., Girardi L., Salasnich B., Dal Cero C., Rubele S. \& Nanni A., 2012, MNRAS, 427, 127 \\
Cayrel R., 1988, in Cayrel de Strobel G., Spite M., eds, The Impact of Very High S/N Spectroscopy on Stellar Physics.
Kluwer Dordrecht, p.345 \\
Chabrier G., 2001, ApJ, 554, 1274 \\
Cutri R. M., Skrutskie M. F., van Dyk S., Beichman C. A., Carpenter J. M., 2003, 2MASS All Sky Catalog of point sources. \\
da Silva L., Girardi L., Pasquini L., Setiawan J., von der L{\"u}he O., de Medeiros J. R., Hatzes A., D{\"o}llinger M. P., Weiss A.,
2006, A\&A, 458, 609 \\
EggenO. J.. 1958a, Observatory, 78, 21 \\
Eggen O. J., 1958b, MNRAS, 118, 65 \\
Eggen O. J., 1958c, MNRAS, 118, 154 \\
Eggen O. J., 1996, AJ, 112, 1595 \\
Famaey B., Jorissen A., Luri X., Mayor M., Udry S., Dejonghe H., Turon C., 2005, A\&A, 430, 165 \\
Famaey B., Siebert A., Jorissen A., 2008, A\&A, 483, 453 \\
Feltzing S., Holmberg J., 2000, A\&A, 357. 153 \\
Fux R., 2001, A\&A, 373, 511 \\
Girardi L., Bressan A., Bertelli G., \& Chiosi C., 2000, A\&AS, 141, 371 \\
Hayden M. R., Bovy J., Holtzman J. A. et al. 2015, ApJ, 808, 132 \\
Hinkle K., Wallace L., Valenti J., \& Harmer D. (ed.) 2000, Visible and Near Infrared Atlas of the Arcturus Spectrum
3727$-$9300 \AA (San Francisco, CA:ASP) \\
Holmberg J., Nordstr\"{o}m, B., Andersen, J. 2007, A\&A, 475, 519 \\
Johnson D. R. H., Soderblom D. R., 1987, AJ, 93, 864 \\
Kurucz R.L., 1998, http://kurucz.harvard.edu/ (online data) \\
Liu F., Chen Y. Q., Zhao G., Han I., Lee B. C., Kim K. M., Zhao Z. S., 2012, MNRAS, 422, 2969 \\
Luck R. E., Heiter U., 2006, AJ, 131, 3069 \\
Luck R. E., Heiter U., 2007, AJ, 133, 2464 \\
McWilliam, A., 1998, AJ, 115, 1640 \\
Mishenina T. V., Pignatari, M., Korotin, S. A., et al., 2013, A\&A, 552, A128 \\
Navarro J. F., Helmi A., Freeman K. C., 2004, ApJ, 601, L43 \\
Nordstr\"{o}m B., Mayor M., Andersen J., Holmberg J., Pont F., Jørgensen B. R., Olsen E. H., Udry S., Mowlavi N., 2004,
A\&A, 418, 989 \\
Pakhomov Y. V., Antipova L. I., Boyarchuk A. A., 2011, Astr. Repts., 55, 256 \\
Quillen A. C., Minchev I., 2005, AJ, 130, 576 \\
Ram{\'{\i}}rez I., Allende Prieto C., 2011, ApJ, 743, 135 \\
Ram{\'{\i}}rez I., Mel\'{e}ndez J., 2005, ApJ, 626, 446 \\
Ramya P., Reddy B. E., Lambert D. L., 2012, MNRAS, 425, 3188 \\
Reddy, A.B.S., Giridhar, S., Lambert, D.L. 2015, MNRAS, 450, 430 \\
Reddy B. E., Lambert D. L., Allende Prieto C., 2006, MNRAS, 367, 1329 \\
Reddy B. E., Tomkin J., Lambert D. L., Allende Prieto C., 2003, MNRAS, 340, 304 \\
Schlegel D. J., Finkbeiner D. P., Davis M., 1998, ApJ, 500, 525 \\
Sellwood J. A., 2014, Rev. Mod. Phys., 86, 1 \\
Sneden C., 1973, PhD Thesis, University of Texas, Austin \\
Soubiran C., Bienaym{\'e} O., Siebert A., 2003, A\&A, 398, 141 \\
Soubiran C., Girard P., 2005, A\&A, 438, 139 \\
Tull R. G., MacQueen P. J., Sneden C., Lambert D. L., 1995, PASP, 107, 251 \\
van Leeuwen F., 2007, A\&A, 474, 653 \\
Zhao J., Zhao G., Chen Y., 2009, ApJ, 692, L113 \\

%\end{thebibliography}

\clearpage

\onecolumn

\begin{center}

\begin{longtable}{cclll}
\caption{Adopted Iron line data} \\ \hline \\
Wavelength &  LEP  &   log $gf$  &  $W_{\lambda_\odot}$   &  log $\varepsilon_{\odot}$  \\
\AA\          &  eV   &    &  m\AA\ &  dex  \\ \\  \hline  \hline 
\endfirsthead
\multicolumn{5}{c} {{ \tablename\ \thetable{} -- continued from previous page}} \\ \hline \\
Wavelength &  LEP  &   log $gf$  &  $W_{\lambda_\odot}$   &  log $\varepsilon_{\odot}$  \\
\AA\    &  eV   &             &    m\AA\       &   dex       \\ \\  \hline  \hline 
\endhead
\hline \hline
\endlastfoot
\multicolumn{5}{c}{Fe I}  \\ \\ 
5295.31 &  4.420 &  -1.590 &  29.1 &  7.53  \\
5358.12 &  3.300 &  -3.162 &  9.5  &  7.43  \\
5379.57 &  3.690 &  -1.510 &  60.5 &  7.39  \\
5386.33 &  4.150 &  -1.670 &  32.1 &  7.43 \\
5441.34 &  4.310 &  -1.630 &  30.3 &  7.49  \\
5638.26 &  4.220 &  -0.770 &  75.8 &  7.41  \\
5661.35 &  4.280 &  -1.756 &  21.7 &  7.38  \\
5679.02 &  4.652 &  -0.750 &  58.2 &  7.45  \\
5705.46 &  4.301 &  -1.355 &  37.7 &  7.36  \\
5731.76 &  4.260 &  -1.200 &  56.5 &  7.52  \\
5778.45 &  2.588 &  -3.440 &  22.2 &  7.45  \\
5793.91 &  4.220 &  -1.619 &  33.2 &  7.45  \\
5849.69 &  3.695 &  -2.930 &  6.6  &  7.38  \\
5855.08 &  4.608 &  -1.478 &  20.9 &  7.38  \\
5856.10 &  4.294 &  -1.558 &  32.5 &  7.44  \\
5858.79 &  4.220 &  -2.180 &  12.5 &  7.44  \\
5859.60 &  4.550 &  -0.608 &  69.3 &  7.41  \\
5905.67 &  4.650 &  -0.690 &  56.8 &  7.36  \\
5927.79 &  4.650 &  -0.990 &  41.5 &  7.38  \\
5929.68 &  4.550 &  -1.310 &  39.5 &  7.57  \\
6003.01 &  3.880 &  -1.060 &  81.7 &  7.48  \\
6027.05 &  4.076 &  -1.090 &  63.2 &  7.35  \\
6056.00 &  4.730 &  -0.400 &  70.6 &  7.37  \\
6079.01 &  4.650 &  -1.020 &  44.9 &  7.47  \\
6093.64 &  4.607 &  -1.300 &  30.2 &  7.42  \\
6096.66 &  3.984 &  -1.810 &  36.9 &  7.48 \\
6151.62 &  2.176 &  -3.282 &  48.7 &  7.42  \\
6159.38 &  4.610 &  -1.830 &  12.6 &  7.45  \\
6165.36 &  4.143 &  -1.460 &  43.9 &  7.42  \\
6187.99 &  3.940 &  -1.620 &  46.1 &  7.43  \\
6240.65 &  2.223 &  -3.287 &  47.8 &  7.45  \\
6270.23 &  2.858 &  -2.540 &  51.6 &  7.40 \\
6271.28 &  3.330 &  -2.703 &  23.8 &  7.45  \\
6436.41 &  4.186 &  -2.360 &  10.0 &  7.46  \\
6518.37 &  2.830 &  -2.450 &  56.2 &  7.36  \\
6581.21 &  1.480 &  -4.680 &  20.8 &  7.51  \\
6591.33 &  4.593 &  -1.950 &  10.4 &  7.44  \\
6608.04 &  2.279 &  -3.914 &  17.2 &  7.43  \\
6699.14 &  4.590 &  -2.100 &  7.9  &  7.45  \\
6703.57 &  2.759 &  -3.023 &  36.1 &  7.46  \\
6705.10 &  4.607 &  -0.980 &  45.4 &  7.38  \\
6713.75 &  4.795 &  -1.400 &  20.8 &  7.44  \\
6725.36 &  4.103 &  -2.167 &  17.0 &  7.44  \\
6726.67 &  4.607 &  -1.030 &  46.0 &  7.43  \\
6733.15 &  4.638 &  -1.400 &  25.9 &  7.43  \\
6739.52 &  1.560 &  -4.794 &  11.5 &  7.39  \\
6793.26 &  4.076 &  -2.326 &  12.4 &  7.41  \\
6810.26 &  4.607 &  -0.986 &  48.6 &  7.44 \\
6828.59 &  4.640 &  -0.820 &  54.6 &  7.41  \\
6837.01 &  4.590 &  -1.687 &  17.6 &  7.44  \\
6842.69 &  4.640 &  -1.220 &  39.1 &  7.52  \\
6843.66 &  4.550 &  -0.830 &  59.5 &  7.42  \\
6857.25 &  4.076 &  -2.038 &  22.3 &  7.43  \\
6971.94 &  3.020 &  -3.340 &  12.6 &  7.41  \\
6999.88 &  4.100 &  -1.460 &  53.8 &  7.52  \\
7022.95 &  4.190 &  -1.150 &  63.6 &  7.47  \\
7132.99 &  4.080 &  -1.650 &  42.1 &  7.47  \\
7751.12 &  4.990 &  -0.730 &  45.3 &  7.43  \\
7802.51 &  5.080 &  -1.310 &  15.4 &  7.41  \\
7807.92 &  4.990 &  -0.509 &  58.8 &  7.44   \\  \\

\multicolumn{5}{c}{Fe II}  \\ \\ 

5234.62 &  3.221 &  -2.180 &  83.3 &  7.36 \\
5264.80 &  3.230 &  -3.130 &  47.9 &  7.53  \\
5414.07 &  3.221 &  -3.580 &  27.5 &  7.50  \\
5425.26 &  3.200 &  -3.220 &  41.3 &  7.44  \\
6149.25 &  3.889 &  -2.630 &  35.7 &  7.36  \\
6247.56 &  3.892 &  -2.271 &  52.8 &  7.39  \\
6369.46 &  2.891 &  -4.110 &  19.4 &  7.48  \\
6432.68 &  2.891 &  -3.570 &  40.9 &  7.46  \\
6456.39 &  3.903 &  -2.065 &  62.3 &  7.39  \\
6516.08 &  2.891 &  -3.310 &  52.6 &  7.45 \\ \hline
\end{longtable}

\end{center}

\clearpage

%\LTcapwidth=\textwidth

{\small  

\begin{center}

\begin{longtable}{lcccccccc}
\caption{Derived Atmospheric Parameters of giants from the  Hercules stream. Columns are self explanatory.} \\ \hline \\
Star  & ($T_{\rm eff}$)$_{J-K}$ & ($T_{\rm eff}$)$_{V-K}$ & ($T_{\rm eff}$)$_{\rm spec}$  & log $g$ (phot) & log $g$ (spec) 
                                                                           & $\xi_{t}$ & [M/H] & Age \\
      &  K        &     K      &    K       &    cm s$^{-2}$  &   cm s$^{-2}$   & km s$^{-1}$ &  dex  & Gyr
\\ \hline \hline

HIP 258    &  4304 &  4953 &  4770 &  2.61 $\pm$ 0.05 &  2.98 &  1.46 &  +0.27 &   0.7 $\pm$ 0.2   \\
HIP 504    &  4833 &  4811 &  4830 &  2.65 $\pm$ 0.08 &  2.80 &  1.50 &  +0.17 &   0.8 $\pm$ 0.3   \\
HIP 3546   &  4617 &  4572 &  4640 &  2.52 $\pm$ 0.12 &  2.58 &  1.55 &  +0.25 &   1.5 $\pm$ 1.3    \\
HIP 3719   &  4752 &  4706 &  4840 &  2.60 $\pm$ 0.09 &  2.76 &  1.53 &  +0.04 &   0.9 $\pm$ 0.4    \\
HIP 4486   &  4138 &  4649 &  4770 &  2.96 $\pm$ 0.07 &  3.08 &  1.25 &  +0.17 &   3.4 $\pm$ 1.0     \\
HIP 6682   &  4858 &  4886 &  4920 &  3.18 $\pm$ 0.07 &  3.32 &  1.22 &  +0.16 &   2.5 $\pm$ 0.7   \\
HIP 7119   &  4699 &  4806 &  4960 &  2.78 $\pm$ 0.22 &  3.00 &  1.46 &  +0.09 &   1.0 $\pm$ 0.6   \\
HIP 7710   &  4486 &  4580 &  4650 &  2.43 $\pm$ 0.07 &  2.65 &  1.35 &  $-$0.11 & 6.1 $\pm$ 2.8   \\
HIP 8926   &  4918 &  4997 &  5100 &  3.02 $\pm$ 0.11 &  3.16 &  1.44 &  +0.23   & 0.8 $\pm$ 0.2   \\
HIP 8984   &  4975 &  4656 &  4800 &  2.71 $\pm$ 0.12 &  2.70 &  1.47 &  +0.05   & 1.8 $\pm$ 0.7   \\
HIP 9307   &  4896 &  5019 &  4920 &  2.64 $\pm$ 0.04 &  2.85 &  1.51 &  +0.01   & 0.7 $\pm$ 0.2  \\
HIP 9517   &  4940 &  4799 &  4800 &  2.58 $\pm$ 0.11 &  2.62 &  1.46 &  $-$0.01 & 2.4 $\pm$ 1.3   \\
HIP 11117  &  4575 &  4757 &  4820 &  3.19 $\pm$ 0.08 &  3.34 &  1.23 &  +0.38  &  2.8 $\pm$ 0.9   \\
HIP 13786  &  4306 &  4693 &  4650 &  2.52 $\pm$ 0.13 &  2.60 &  1.62 &  +0.15  &  3.9 $\pm$ 2.5 \\
HIP 18865  &  4737 &  4555 &  4960 &  3.05 $\pm$ 0.25 &  3.10 &  1.42 &  +0.30  &  1.4 $\pm$ 0.6  \\
HIP 19287  &  4552 &  4501 &  4690 &  2.72 $\pm$ 0.10 &  2.65 &  1.54 &  +0.24  &  1.8 $\pm$ 0.6   \\
HIP 20540  &  4797 &  5078 &  4965 &  2.84 $\pm$ 0.06 &  3.20 &  1.26 &  +0.03  &  1.3 $\pm$ 0.2  \\
HIP 20771  &  4346 &  4966 &  4870 &  2.78 $\pm$ 0.05 &  3.00 &  1.42 &  +0.31  &  1.2 $\pm$ 0.3  \\
HIP 22176  &  4460 &  4489 &  4660 &  2.71 $\pm$ 0.08 &  2.90 &  1.51 &  +0.30  &  1.7 $\pm$ 0.6  \\
HIP 22661  &  4488 &  4389 &  4630 &  2.58 $\pm$ 0.10 &  2.77 &  1.70 &  +0.32  &  1.8 $\pm$ 1.1   \\
HIP 22765  &  4631 &  4710 &  4730 &  2.62 $\pm$ 0.09 &  2.75 &  1.59 &  +0.20  &  1.3 $\pm$ 0.7   \\
HIP 28168  &  4455 &  4511 &  4670 &  2.87 $\pm$ 0.08 &  3.15 &  1.37 &  +0.40  &  2.5 $\pm$ 0.8   \\
HIP 28417  &  4770 &  4758 &  4900 &  2.68 $\pm$ 0.12 &  2.75 &  1.51 &  $-$0.09 & 1.7 $\pm$ 0.7    \\
HIP 28677  &  5323 &  4981 &  4910 &  3.04 $\pm$ 0.09 &  3.10 &  1.30 &  $-$0.06 & 3.6 $\pm$ 1.0    \\
HIP 29949  &  4510 &  4540 &  4710 &  2.55 $\pm$ 0.09 &  2.70 &  1.53 &  +0.14 &   2.7 $\pm$ 1.3   \\
HIP 31039  &  5035 &  5170 &  5030 &  3.15 $\pm$ 0.06 &  3.35 &  1.25 &  -0.05 &   2.1 $\pm$ 0.5   \\
HIP 32261  &  4965 &  4747 &  4850 &  3.05 $\pm$ 0.05 &  3.23 &  1.26 &  +0.32 &   1.3 $\pm$ 0.2    \\
HIP 32844  &  4561 &  4329 &  4480 &  2.12 $\pm$ 0.07 &  2.34 &  1.61 &  +0.17 &   0.7 $\pm$ 0.2    \\
HIP 35146  &  4392 &  4747 &  4790 &  2.80 $\pm$ 0.04 &  3.06 &  1.29 &  +0.26 &   1.4 $\pm$ 0.2    \\
HIP 36647  &  4766 &  4649 &  4680 &  2.66 $\pm$ 0.14 &  2.65 &  1.45 &  +0.10 &   3.5 $\pm$ 1.7     \\
HIP 37049  &  4895 &  4820 &  4930 &  2.77 $\pm$ 0.07 &  3.00 &  1.48 &  +0.10 &   1.1 $\pm$ 0.3    \\
HIP 37441  &  4683 &  4745 &  4930 &  2.62 $\pm$ 0.11 &  2.75 &  1.62 &  $-$0.16 & 1.7 $\pm$ 0.9    \\
HIP 48140  &  4489 &  4549 &  4680 &  2.60 $\pm$ 0.10 &  2.65 &  1.46 &  +0.27 &   1.4 $\pm$ 0.8     \\
HIP 48417  &  4493 &  4324 &  4440 &  2.44 $\pm$ 0.11 &  2.33 &  1.41 &  +0.05 &   7.1 $\pm$ 2.6    \\
HIP 50526  &  4567 &  4820 &  4870 &  2.71 $\pm$ 0.08 &  2.78 &  1.52 &  +0.06 &   1.2 $\pm$ 0.5   \\
HIP 51047  &  4716 &  4673 &  4830 &  2.76 $\pm$ 0.07 &  2.80 &  1.52 &  +0.12 &   1.4 $\pm$ 0.4    \\
HIP 52882  &  4780 &  4726 &  4670 &  2.29 $\pm$ 0.08 &  2.58 &  1.51 &  $-$0.02 & 0.9 $\pm$ 0.4    \\
HIP 58654  &  4854 &  4564 &  4840 &  2.61 $\pm$ 0.09 &  2.64 &  1.47 &  $-$0.02 & 1.4 $\pm$ 0.7   \\
HIP 87629  &  5113 &  5013 &  5090 &  3.26 $\pm$ 0.08 &  3.42 &  1.27 &  +0.08 &   1.6 $\pm$ 0.3    \\
HIP 94576  &  4949 &  4884 &  5050 &  2.92 $\pm$ 0.08 &  3.20 &  1.45 &  +0.28 &   0.8 $\pm$ 0.2    \\
HIP 95375  &  4736 &  4786 &  4760 &  2.83 $\pm$ 0.06 &  3.00 &  1.31 &  +0.15 &   1.8 $\pm$ 0.6     \\
HIP 96028  &  4535 &  4581 &  4700 &  2.76 $\pm$ 0.07 &  2.93 &  1.29 &  +0.16 &   2.2 $\pm$ 0.9    \\
HIP 96294  &  4601 &  4823 &  4880 &  2.65 $\pm$ 0.09 &  3.00 &  1.48 &  +0.30 &   0.7 $\pm$ 0.2    \\
HIP 102010 &  4376 &  4720 &  4710 &  2.51 $\pm$ 0.14 &  2.70 &  1.56 &  +0.13 &   1.4 $\pm$ 1.3    \\
HIP 104035 &  4871 &  4966 &  5040 &  2.85 $\pm$ 0.09 &  3.05 &  1.42 &  +0.20 &   1.0 $\pm$ 0.2    \\
HIP 105502 &  4820 &  5021 &  4770 &  2.62 $\pm$ 0.04 &  2.86 &  1.55 &  +0.17 &   0.8 $\pm$ 0.2    \\
HIP 106551 &  4464 &  4901 &  4840 &  2.68 $\pm$ 0.04 &  2.88 &  1.59 &  +0.18 &   0.9 $\pm$ 0.2    \\
HIP 107502 &  4380 &  4865 &  4610 &  2.51 $\pm$ 0.08 &  2.75 &  1.38 &  +0.25 &   2.9 $\pm$ 1.5     \\
HIP 108012 &  4348 &  4459 &  4580 &  2.43 $\pm$ 0.08 &  2.65 &  1.44 &  +0.14 &   3.4 $\pm$ 1.9    \\
HIP 108914 &  4806 &  4785 &  4860 &  2.79 $\pm$ 0.05 &  3.15 &  1.44 &  +0.42 &   1.2 $\pm$ 0.3   \\
HIP 109387 &  4864 &  5011 &  5000 &  3.20 $\pm$ 0.06 &  3.40 &  1.27 &  +0.20 &   1.5 $\pm$ 0.3    \\
HIP 109585 &  4233 &  5249 &  4760 &  2.65 $\pm$ 0.05 &  2.72 &  1.52 &  +0.18 &   1.0 $\pm$ 0.3    \\
HIP 111728 &  4407 &  4569 &  4610 &  2.69 $\pm$ 0.10 &  2.94 &  1.47 &  +0.33 &   2.4 $\pm$ 1.1    \\
HIP 113144 &  4735 &  4762 &  4980 &  3.23 $\pm$ 0.07 &  3.40 &  1.19 &  +0.10 &   2.3 $\pm$ 0.6     \\
HIP 114742 &  ...  &  ...  &  4750 &  2.49 $\pm$ 0.06 &  2.93 &  1.63 &  +0.35 &   0.5 $\pm$ 0.1     \\
HIP 115899 &  5091 &  4840 &  4910 &  2.70 $\pm$ 0.06 &  2.95 &  1.44 &  +0.12 &   0.9 $\pm$ 0.3     \\
HIP 116348 &  4714 &  4833 &  4830 &  2.72 $\pm$ 0.09 &  2.77 &  1.53 &  +0.10 &   1.4 $\pm$ 0.5     \\
HIP 116644 &  4605 &  4553 &  4760 &  2.60 $\pm$ 0.15 &  2.88 &  1.45 &  +0.13 &   1.2 $\pm$ 0.9   \\  \hline
\end{longtable}
\end{center}
}

\clearpage

\begin{center}
\begin{longtable}{lclllc}
\caption{Adopted line list for elements other than Iron} \\ \hline \\  
Species & Wavelength &  LEP  &   log$gf$  &  $W_{\lambda_\odot}$   &  log $\varepsilon_{\odot}$   \\
     &  \AA\   &  eV   &     &    m\AA  &   dex  \\ \\  \hline  \hline  
\endfirsthead
\multicolumn{6}{c} {{ \tablename\ \thetable{} -- continued from previous page}} \\ \hline \\
Species & Wavelength  &  LEP  &   log$gf$  &  $W_{\lambda_\odot}$   &  log $\varepsilon_{\odot}$   \\
        &  \AA\          &  eV   &             &    m\AA\       &   dex    \\ \\  \hline  \hline 
\endhead
\hline \hline
\endlastfoot
{}Na I               & 6154.23  &  2.10 &  -1.55 &  36.6 &  6.28   \\
                  & 6160.75  &  2.10 &  -1.25 &  56.5 &  6.29    \\
Mg I               & 5711.09  &  4.34 &  -1.73 &  104.1 &  7.54   \\
                  & 6318.72  &  5.11 &  -1.95 &  44.5 &  7.58 \\
                  & 6319.24  &  5.11 &  -2.32 &  27.4 &  7.65  \\
                  & 7657.61  &  5.11 &  -1.28 &  98.5 &  7.59 \\
Al I               & 6696.02  &  3.14 &  -1.48 &  36.9 &  6.40  \\
                  & 6698.67  &  3.14 &  -1.78 &  20.8 &  6.36  \\
                  & 7835.31  &  4.02 &  -0.69 &  41.1 &  6.41  \\
                  & 7836.13  &  4.02 &  -0.45 &  55.0 &  6.36  \\
Si I               & 5690.42  &  4.93 &  -1.77 &  48.1 &  7.47  \\
                  & 5701.10  &  4.93 &  -1.95 &  37.9 &  7.47  \\
                  & 5772.15  &  5.08 &  -1.65 &  52.3 &  7.55  \\
                  & 6142.49  &  5.62 &  -1.540 &  33.3 &  7.58  \\
                  & 6145.02  &  5.61 &  -1.479 &  37.6 &  7.59  \\
Ca I               & 5260.39  &  2.52 &  -1.72 &  32.1 &  6.27  \\
                  & 5867.56  &  2.93 &  -1.57 &  22.6 &  6.27  \\
                  & 6166.44  &  2.52 &  -1.14 &  69.1 &  6.31  \\
                  & 6169.04  &  2.52 &  -0.80 &  90.3 &  6.32  \\
                  & 6169.56  &  2.53 &  -0.48 &  108.7 &  6.28  \\
                  & 6455.60  &  2.52 &  -1.34 &  55.9 &  6.28  \\
                  & 6471.66  &  2.53 &  -0.69 &  90.6 &  6.20  \\
                  & 6499.65  &  2.52 &  -0.82 &  84.7 &  6.23  \\
Sc II              & 5357.20  &  1.51 &  -2.11 &  4.8 &  3.19  \\
                  & 5552.23  &  1.46 &  -2.28 &  4.6 &  3.28  \\
                  & 5684.21  &  1.51 &  -1.07 &  37.1 &  3.24  \\
                  & 6245.64  &  1.51 &  -1.04 &  35.2 &  3.14  \\
                  & 6300.75  &  1.51 &  -1.95 &  8.2 &  3.24 \\
                  & 6320.84  &  1.50 &  -1.92 &  8.9 &  3.24  \\
Ti I               & 5295.77  &  1.07 &  -1.58 &  13.2 &  4.95  \\
                  & 5490.15  &  1.46 &  -0.88 &  22.0 &  4.89  \\
                  & 5702.66  &  2.29 &  -0.59 &  7.3 &  4.83 \\
                  & 5716.44  &  2.30 &  -0.72 &  5.9 &  4.87  \\
                  & 6092.79  &  1.89 &  -1.32 &  4.1 &  4.89  \\
                  & 6303.75  &  1.44 &  -1.51 &  8.8 &  4.99  \\
                  & 6312.23  &  1.46 &  -1.50 &  8.1 &  4.95 \\
                  & 6599.10  &  0.90 &  -2.03 &  9.3 &  4.98  \\
                  & 7357.73  &  1.44 &  -1.07 &  22.2 &  4.97  \\
Ti II              & 4583.41  &  1.17 &  -2.87 &  33.0 &  5.06 \\
                  & 4708.66  &  1.24 &  -2.37 &  53.3 &  5.04  \\ 
                  & 5336.78  &  1.58 &  -1.63 &  71.7 &  4.96  \\
                  & 5418.77  &  1.58 &  -2.11 &  48.5 &  4.97 \\
V I                & 6039.73  &  1.06 &  -0.65 &  12.3 &  3.93  \\
                  & 6081.44  &  1.05 &  -0.58 &  13.0 &  3.87  \\
                  & 6090.21  &  1.08 &  -0.06 &  32.3 &  3.89  \\
                  & 6119.53  &  1.06 &  -0.32 &  21.0 &  3.88  \\
                  & 6135.36  &  1.05 &  -0.75 &  10.2 &  3.91  \\
                  & 6274.65  &  0.27 &  -1.67 &  6.9 &  3.87  \\
Cr I               & 5287.20  &  3.44 &  -0.89 &  11.3 &  5.65  \\
                  & 5300.74  &  0.98 &  -2.08 &  59.3 &  5.60  \\
                  & 5304.18  &  3.46 &  -0.68 &  16.0 &  5.63 \\
                  & 5628.62  &  3.42 &  -0.76 &  14.7 &  5.62  \\
                  & 5781.16  &  3.01 &  -1.00 &  16.7 &  5.54  \\
                  & 6882.48  &  3.44 &  -0.38 &  32.4 &  5.67  \\
                  & 6883.00  &  3.44 &  -0.42 &  30.5 &  5.67  \\
Mn I               & 4671.69  &  2.89 &  -1.66 &  14.8 &  5.47  \\
                  & 4739.11  &  2.94 &  -0.60 &  60.7 &  5.39  \\
                  & 5004.89  &  2.92 &  -1.64 &  14.0 &  5.45  \\
Co I               & 5280.63  &  3.63 &  -0.03 &  20.3 &  4.90 \\
                  & 5352.04  &  3.58 &   0.06 &  25.1 &  4.88  \\
                  & 5647.23  &  2.28 &  -1.56 &  13.9 &  4.90  \\
                  & 6455.00  &  3.63 &  -0.25 &  14.8 &  4.89  \\
Ni I             & 5088.96  &  3.678 &  -1.240 &  28.2 &  6.19  \\
                  & 5094.42  &  3.833 &  -1.074 &  30.4 &  6.22  \\
                  & 5115.40  &  3.834 &  -0.281 &  75.2 &  6.29  \\
                  & 6111.08  &  4.088 &  -0.808 &  33.6 &  6.22  \\
                  & 6130.14  &  4.266 &  -0.938 &  21.6 &  6.24  \\
                  & 6175.37  &  4.089 &  -0.550 &  47.7 &  6.24  \\
                    & 6176.80  &  4.09 &  -0.26 &  62.0 &  6.22  \\
                     & 6177.25  &  1.826 &  -3.508 &  14.1 &  6.22  \\
                  & 6772.32  &  3.658 &  -0.972 &  48.0 &  6.24  \\
                  & 7797.59  &  3.900 &  -0.348 &  75.3 &  6.27  \\
                  & 7826.77  &  3.700 &  -1.840 &  12.5 &  6.23  \\
Zn I               & 4810.54  &  4.080 &  -0.170 &  71.6 &  4.45  \\
                  & 6362.35  &  5.790 &   0.140 &  21.2 &  4.53  \\
Ba II              & 5853.68  &  0.604 &  -1.000 &  60.3 &  2.15  \\
                  & 6141.73  &  0.704 &  -0.032 & 109.1 &  2.23  \\

\end{longtable}
\end{center}

\clearpage

\begin{table}
% \centering
% \begin{minipage}{140mm}
  \caption{Derived solar abundances compared with the values of Asplund et al. (2009). $\sigma$ is the standard 
  deviation or line to line scatter and N is the number of lines used}
  \begin{center}
  \begin{tabular}{@{}lcccr@{}}
  \hline \\

Species   & (log$\varepsilon\pm\sigma$) & N   & (log$\varepsilon\pm\sigma$)  &  Difference \\
          &   dex                       &      &  dex                        &  dex  \\  
          &  (Current)                  &      & (Asplund)                &      \\ \\ \hline \hline \\
{}Na I   & 6.29 $\pm$ 0.01 & 2   & 6.24 $\pm$ 0.04  & $+$0.05    \\
Mg I   & 7.59 $\pm$ 0.04 & 4   & 7.60 $\pm$ 0.04  & $-$0.01  \\
Al I   & 6.38 $\pm$ 0.03 & 4   & 6.45 $\pm$ 0.03  & $-$0.07   \\
Si I   & 7.53 $\pm$ 0.06 & 5   & 7.51 $\pm$ 0.03  & $+$0.02   \\
Ca I   & 6.27 $\pm$ 0.04 & 8   & 6.34 $\pm$ 0.04  & $-$0.07   \\
Sc II  & 3.21 $\pm$ 0.05 & 6  & 3.15 $\pm$ 0.04  & $+$0.06  \\
Ti I   & 4.93 $\pm$ 0.06 & 9   & 4.95 $\pm$ 0.05  & $-$0.02   \\
Ti II  & 5.01 $\pm$ 0.05 & 4   & 4.95 $\pm$ 0.05  & $+$0.06    \\
V I    & 3.89 $\pm$ 0.03 & 6   & 3.93 $\pm$ 0.08  & $-$0.04   \\
Cr I   & 5.63 $\pm$ 0.05 & 7   & 5.64 $\pm$ 0.04  & $-$0.01  \\
Mn I   & 5.44 $\pm$ 0.04 & 3    & 5.43 $\pm$ 0.04  & $+$0.01   \\
Fe I   & 7.44 $\pm$ 0.05 & 60  & 7.50 $\pm$ 0.04  & $-$0.06   \\
Fe II  & 7.44 $\pm$ 0.06 & 10  & 7.50 $\pm$ 0.04  & $-$0.06  \\
Co I   & 4.85 $\pm$ 0.02 & 4   & 4.99 $\pm$ 0.07 & $-$0.14    \\
Ni I   & 6.23 $\pm$ 0.03 & 11  & 6.22 $\pm$ 0.04  & $+$0.01    \\
Zn I   & 4.49 $\pm$ 0.05 & 1   & 4.56 $\pm$ 0.05  & $-$0.06   \\
Ba II  & 2.10 $\pm$ 0.02 & 2   & 2.18 $\pm$ 0.09 & $-$0.08   \\ \\
\hline \hline

\end{tabular}
%\end{minipage}
\end{center}
\end{table}

\clearpage

\begin{landscape}

\begin{longtable}{lrrrrlrrr}
\caption{The elemental abundances for Hercules stream members for the elements Na, Mg, Al, Si, Ca and Sc given as [El/Fe]. 
$\sigma$ is the standard deviation or line to line scatter where N is the number of lines. Hyperfine 
corrections are applied for  Sc. Mean of [FeI/H] and [FeII/H] is  used as [Fe/H] throughout}  \\ \hline \\
Star&[FeI/H] $\sigma$ N&[FeII/H]  $\sigma$ N&[Na/Fe] $\sigma$ N&[Mg/Fe]
 $\sigma$ N&[Al/Fe]  $\sigma$ N&[Si/Fe] $\sigma$ N&[Ca/Fe] $\sigma$ N & [Sc/Fe] $\sigma$ N  \\
\\  \hline  \hline 
\endfirsthead
\multicolumn{9}{c} {{ \tablename\ \thetable{} -- continued from previous page}} \\
\\ \hline \\
Star&[FeI/H] $\sigma$ N&[FeII/H] $\sigma$ N&[Na/Fe] $\sigma$ N&[Mg/Fe]
$\sigma$ N&[Al/Fe] $\sigma$ N&[Si/Fe] $\sigma$ N&[Ca/Fe] $\sigma$ N & [Sc/Fe] $\sigma$ N  \\
\\ \\  \hline  \hline 
\endhead
\hline \hline
\endlastfoot

{}HIP 258  & +0.26  0.08  56 & +0.27  0.05  8   & 0.25  0.01  2  & -0.04  0.04  4    & 0.18  0.06  4    & 0.09  0.06  5  & -0.05  0.06  8	    & 0.06   0.04   6     \\      	   
HIP 504  & +0.17  0.07  57 & +0.16  0.03  8   & 0.09  0.04  2  & 0.02  0.03  4     & 0.12  0.05  4     & 0.06  0.06  5  & 0.03  0.06  8	    & -0.03  0.05   6    \\        	   
HIP 3546  & +0.24  0.07  55 & +0.24  0.06  8   & 0.27  0.02  2  & 0.00  0.04  4     & 0.13  0.03  4     & 0.11  0.07  5  & -0.03  0.07  8    & -0.06   0.05   6     \\      	 
HIP 3719  & +0.04  0.06  57 & +0.03  0.06  8   & 0.07  0.01  2  & 0.07  0.02  3     & 0.12  0.06  4     & 0.08  0.06  5  & 0.06  0.04  8	    & 0.04   0.03   6     \\      	   
HIP 4486  & +0.16  0.08  56 & +0.15  0.06  8   & 0.09  0.03  2  & 0.01  0.03  4     & 0.16  0.07  4      & 0.07  0.07  5  & 0.02  0.05  8    & 0.06   0.02   6     \\      	       
HIP 6682  & +0.13  0.06  57 & +0.15  0.04  8   & 0.00  0.02  2  & -0.02  0.01  4    & 0.06  0.02  4     & 0.03  0.05  5  & 0.01  0.07  8	    & 0.03    0.05   6      \\      	  
HIP 7119  & +0.07  0.06  57 & +0.07  0.05  8   & 0.09  0.01  2  & -0.01  0.06  4    & 0.03  0.03  4    & 0.05  0.06  5  & 0.05  0.05  8	    & 0.03    0.04   6      \\       	 
HIP 7710  & -0.13  0.06  57 & -0.13  0.05  8   & 0.03  0.05  2  & 0.11  0.03  4     & 0.21  0.04  4    & 0.12  0.06  5  & 0.08  0.04  8	    & 0.06    0.02   6      \\      	
HIP 8926  & +0.22  0.06  57 & +0.21  0.04  8   & 0.16  0.02  2  & -0.06  0.02  4    & 0.04  0.04  4    & 0.04  0.06  5  & 0.06  0.05  8	    & -0.03   0.03   6     \\     	  
HIP 8984  & +0.04  0.06  57 & +0.03  0.03  8   & 0.07  0.03  2  & 0.05  0.01  4     & 0.13  0.03  4    & 0.10  0.06  5  & 0.03  0.05  8	    & 0.04   0.05   6     \\     	 
HIP 9307  & +0.01  0.06  56 & +0.01  0.03  8   & 0.07  0.03  2  & 0.04  0.05  4     & 0.12  0.06  4    & 0.05  0.06  5  & 0.05  0.05  8	    & 0.04    0.05   6      \\      	 
HIP 9517  & +0.00  0.06  57 & -0.02  0.07  8   & 0.02  0.03  2  & 0.05  0.05  3     & 0.23  0.07  4    & 0.07  0.05  5  & 0.10  0.04  8	    & -0.10   0.05   4     \\      	  
HIP 11117  & +0.38  0.08  56 & +0.37  0.07  8   & 0.22  0.02  2   & 0.04  0.05  4    & 0.17  0.04  4    & 0.05  0.06  5  & -0.05  0.07  8    & 0.03   0.04   6     \\             
HIP 13786  & +0.13  0.08  55 & +0.14  0.07  9   & 0.24  0.06  2  & -0.06  0.07  4    & 0.30  0.03  4    & 0.09  0.07  5  & 0.04  0.06  8     & 0.02   0.09   6     \\            
HIP 18865  & +0.33  0.07  56 & +0.33  0.07  8   & 0.29  0.01  2  & -0.03  0.01  2    & 0.06  0.05  4    & 0.08  0.07  5  & 0.02  0.05  8     & 0.02    0.08   6      \\            
HIP 19287  & +0.24  0.07  57 & +0.22  0.07  8   & 0.17  0.00  2  & 0.01  0.03  4     & 0.12  0.03  4    & 0.06  0.07  5  & 0.00  0.07  8     & -0.03   0.04   4     \\            
HIP 20540  & +0.03  0.06  57 & +0.03  0.05  8   & 0.12  0.02  2  & 0.00  0.03  3     & 0.09  0.02  4    & 0.04  0.06  5  & 0.07  0.04  8     & 0.02    0.05   4      \\            
HIP 20771  & +0.29  0.07  57 & +0.30  0.06  8   & 0.17  0.00  2  & -0.12  0.05  3    & 0.06  0.05  4    & 0.03  0.05  5  & -0.01  0.05  8    & -0.04  0.05   6    \\            
HIP 22176  & +0.32  0.07  55 & +0.31  0.07  8   & 0.40  0.03  2  & -0.05  0.03  3    & 0.14  0.02  4    & 0.06  0.06  5  & -0.02  0.07  8    & 0.07   0.04   5     \\            
HIP 22661  & +0.31  0.08  55 & +0.33  0.08  8   & 0.38  0.04  2  & 0.04  0.07  4     & 0.20  0.05  4    & 0.14  0.07  5  & -0.12  0.08  7    & 0.03    0.04   6      \\            
HIP 22765  & +0.19  0.08  56 & +0.20  0.08  8   & 0.15  0.01  2  & -0.05  0.08  4    & 0.12  0.02  4    & 0.10  0.06  5  & -0.03  0.07  8    & -0.05  0.04   6    \\           
HIP 28168  & +0.42  0.08  57 & +0.44  0.08  8   & 0.18  0.01  2  & -0.07  0.06  4    & 0.13  0.07  4    & 0.02  0.07  5  & -0.11  0.07  8    & 0.08    0.05   6      \\           
HIP 28417  & -0.10  0.06  57 & -0.11  0.05  8   & 0.02  0.02  2  & 0.11  0.05  3     & 0.16  0.05  4    & 0.10  0.08  5  & 0.08  0.04  8     & 0.09   0.05   4     \\          
HIP 28677  & -0.08  0.05  57 & -0.08  0.05  9   & 0.05  0.02  2  & 0.07  0.06  4     & 0.15  0.05  4    & 0.06  0.05  5  & 0.07  0.03  8     & -0.01  0.04   6    \\         
HIP 29949  & +0.13  0.07  57 & +0.12  0.06  8   & 0.17  0.00  2  & 0.17  0.04  3     & 0.22  0.03  4    & 0.12  0.07  5  & 0.05  0.07  8     & 0.02   0.05   6     \\          
HIP 31039  & -0.03  0.06  57 & -0.03  0.04  8   & 0.04  0.04  2  & 0.00  0.05  3     & 0.08  0.02  4    & 0.05  0.05  5  & 0.06  0.03  8     & -0.02   0.06   6     \\          
HIP 32261  & +0.31  0.08  57 & +0.30  0.05  8   & 0.25  0.01  2  & -0.01  0.02  4    & 0.15  0.06  4    & 0.06  0.06  5  & -0.02  0.06  8    & 0.05   0.05   6     \\        
HIP 32844  & +0.17  0.07  57 & +0.19  0.05  8   & 0.10  0.01  2  & -0.04  0.05  4    & 0.12  0.05  4     & 0.10  0.07  5  & -0.10  0.06  8   & 0.01    0.03   5      \\         
HIP 35146  & +0.27  0.07  57 & +0.28  0.06  8   & 0.06  0.04  2  & -0.04  0.06  4    & 0.08  0.06  4     & 0.02  0.06  5  & -0.06  0.06  8   & -0.04  0.04   6    \\         
HIP 36647  & +0.11  0.08  58 & +0.11  0.07  8   & 0.08  0.00  2  & 0.13  0.05  4     & 0.15  0.03  4     & 0.12  0.07  5  & 0.00  0.06  8    & 0.01    0.07   5      \\          
HIP 37049  & +0.09  0.05  57 & +0.09  0.04  8   & 0.11  0.02  2  & -0.04  0.07  3    & 0.09  0.03  4     & 0.05  0.06  5  & 0.02  0.05  8    & 0.03    0.04   6      \\          
HIP 37441  & -0.16  0.05  57 & -0.17  0.03  8   & 0.10  0.04  2  & 0.07  0.07  3     & 0.14  0.03  4     & 0.11  0.06  5  & 0.09  0.04  8    & 0.09   0.04   6     \\        
HIP 48140  & +0.29  0.07  56 & +0.31  0.05  8   & 0.21  0.00  2  & 0.01  0.07  4     & 0.12  0.07  4     & 0.02  0.05  5  & -0.08  0.07  8   & -0.07   0.05   6     \\         
HIP 48417  & +0.03  0.07  57 & +0.05  0.07  8   & 0.12  0.00  2  & 0.05  0.05  3     & 0.19  0.06  4     & 0.09  0.03  5  & -0.01  0.05  8   & 0.03    0.06   5      \\         
HIP 50526  & +0.06  0.06  57 & +0.05  0.04  8   & 0.08  0.02  2  & 0.01  0.03  4     & 0.11  0.04  4     & 0.09  0.06  5  & 0.05  0.05  8    & -0.03  0.05   6    \\         
HIP 51047  & +0.12  0.06  57 & +0.10  0.05  8   & 0.12  0.03  2  & 0.02  0.07  4     & 0.15  0.06  4     & 0.05  0.06  5  & 0.03  0.05  8    & 0.00    0.05   6      \\       
HIP 52882  & -0.02  0.06  57 & -0.04  0.06  8   & 0.17  0.04  2  & 0.03  0.06  4     & 0.11  0.02  4     & 0.06  0.05  5  & 0.02  0.05  8    & 0.06    0.05   6      \\       
HIP 58654  & -0.01  0.06  57 & -0.03  0.03  9   & 0.10  0.03  2  & 0.07  0.02  4     & 0.18  0.03  4     & 0.07  0.05  5  & 0.09  0.06  8    & 0.00    0.04   6      \\       
HIP 87629  & +0.06  0.05  56 & +0.05  0.07  8  & -0.02  0.02  2  & -0.02  0.07  2    & 0.01  0.03  4    & 0.03  0.06  5  & 0.05  0.05  8     & 0.02   0.09   5     \\    
HIP 94576  & +0.27  0.07  54 & +0.26  0.06  8  & 0.44  0.01   2  & 0.10  0.08  2     & 0.21  0.06  4    & 0.03  0.07  5  & 0.06  0.06  8     & 0.04   0.07   6     \\     
HIP 95375  & +0.14  0.06  56 & +0.12  0.06  8  & 0.09  0.03   2  & 0.06  0.07  2     & 0.08  0.04  4    & 0.06  0.07  5  & -0.02  0.06  8    & 0.02    0.07   6      \\     
HIP 96028  & +0.16  0.07  55 & +0.18  0.06  8  & 0.05  0.04   2  & 0.00  0.02  3     & 0.10  0.04  4    & 0.06  0.04  5  & -0.01  0.05  8    & 0.04    0.06   6      \\     
HIP 96294 & +0.29 0.07  55 & +0.27  0.07  8  & 0.23  0.01   2  & -0.06  0.02  3    & 0.06  0.03  4    & 0.02  0.07  5  & 0.00  0.07  8     & 0.06    0.07   6      \\     
HIP 102010  & +0.12  0.07  57 & +0.11  0.05  7   & 0.11  0.05  2  & 0.18  0.04  3     & 0.20  0.05  4     & 0.13  0.06  5 & -0.01  0.08  8   & 0.07   0.06   6     \\   
HIP 104035  & +0.18  0.06  56 & +0.16  0.03  8   & 0.07  0.03  2  & -0.02  0.05  3    & 0.00  0.02  4     & 0.02  0.06  5 & 0.08  0.06  8    & -0.03   0.06   6     \\   
HIP 105502  & +0.18  0.06  56 & +0.18  0.06  7   & -0.01 0.09  2  & 0.03  0.05  2     & 0.13  0.07  4     & 0.06  0.05  5 & -0.03  0.08  8   & 0.09    0.06   4      \\   
HIP 106551  & +0.18  0.08  56 & +0.19  0.06  8   & 0.17  0.01  2  & -0.08  0.02  2    & 0.11  0.07  4     & 0.06  0.05  5 & -0.04  0.07  8   & 0.07   0.08   5     \\   
HIP 107502  & +0.23  0.06  57 & +0.23  0.05  8      & 0.11  0.03  2 & 0.06  0.02  3 & 0.12  0.06  4    & 0.06  0.07  5  & -0.02  0.06  8   & 0.02   0.01   6     \\     
HIP 108012  & +0.16  0.07  56 & +0.15  0.07  8   & 0.20  0.02  2   & 0.08  0.06  3    & 0.12  0.05  4    & 0.10  0.06  5  & -0.01  0.07  8   & 0.01   0.04   6     \\          
HIP 108914  & +0.43  0.08  55 & +0.42  0.06  8   & 0.29  0.05  2  & -0.08  0.05  2    & 0.14  0.06  4     & 0.07  0.06  5 & -0.06  0.07  8   & 0.08   0.06   6     \\   
HIP 109387  & +0.21  0.07  56 & +0.21  0.06  9   & 0.07  0.01  2  & 0.06  0.06  4     & 0.07  0.03  4     & 0.03  0.04  5 & -0.04  0.05  8   & -0.01   0.05   6     \\   
HIP 109585  & +0.21  0.07  55 & +0.19  0.04  8   & 0.15  0.04  2  & 0.10  0.06  4     & 0.14  0.04  4     & 0.07  0.05  5 & 0.01  0.08  8    & 0.01    0.05   6      \\   
HIP 111728  & +0.31  0.08  55 & +0.32  0.06  8   & 0.34  0.03  2   & 0.07  0.04  3    & 0.22  0.04  4    & 0.10  0.03  5  & -0.06  0.07  8   & 0.08   0.05   6     \\         
HIP 113144  & +0.09  0.07  55 & +0.10  0.05  8   & -0.01  0.02  2 & 0.05  0.10  3     & 0.09  0.06  4     & 0.03  0.04  5 & 0.04  0.05  8    & 0.09   0.05   5     \\   
HIP 114742  & +0.36  0.07  55 & +0.37  0.07  8   & 0.33  0.03  2   & 0.04  0.05  4    & 0.21  0.06  4    & 0.13  0.08  5  & -0.06  0.08  8   & 0.10   0.05   6     \\          
HIP 115899  & +0.09  0.06  56 & +0.11  0.05  8   & 0.04  0.05  2  & 0.01  0.03  3     & 0.03  0.03  4     & 0.06  0.06  5 & 0.04  0.04  8    & 0.05    0.06   6      \\   
HIP 116348  & +0.08  0.06  57 & +0.08  0.05  8   & 0.06  0.02  2   & 0.03  0.05  4    & 0.15  0.04  4    & 0.08  0.06  5  & 0.03  0.05  8    & 0.02    0.04   6      \\          
{}HIP 116644  & +0.13  0.07  56 & +0.15  0.05  8   & 0.09  0.02  2  & -0.01  0.08  4    & 0.00  0.02  4     & 0.07  0.02  5 & -0.05  0.05  8   & 0.02    0.05   6      \\

\end{longtable}

\clearpage

\begin{longtable}{lrrrrrrrrr}
\caption{The elemental abundances for Hercules stream members for the elements Ti, V, Cr, Mn, Co, Ni,  Zn and Ba given as [El/Fe]. 
$\sigma$ is the standard deviation or line to line scatter where N is the number of lines. Hyperfine correction  are
applied for V, Mn, Co and Ba} \\ \hline \\
Star&[TiI/Fe] $\sigma$N&[TiII/Fe] $\sigma$ N&[V/Fe] $\sigma$ N& [Cr/Fe] $\sigma$ N&[Mn/Fe]
$\sigma$ N&[Co/Fe] $\sigma$ N&[Ni/Fe] $\sigma$ N&[Zn/Fe] $\sigma$ N&[Ba/Fe] $\sigma$ N  \\
\\  \hline  \hline 
\endfirsthead
\multicolumn{10}{c} {{ \tablename\ \thetable{} -- continued from previous page}} \\
\\ \hline \\
Star&[TiI/Fe] $\sigma$N&[TiII/Fe] $\sigma$ N&[V/Fe] $\sigma$ N& [Cr/Fe] $\sigma$ N&[Mn/Fe]
$\sigma$ N&[Co/Fe] $\sigma$ N&[Ni/Fe] $\sigma$ N&[Zn/Fe] $\sigma$ N&[Ba/Fe] $\sigma$ N  \\
\\ \\  \hline  \hline 
\endhead
\hline \hline
\endlastfoot

{}HIP 258    & -0.05  0.08  8   & -0.01  0.04  4  & -0.03  0.05  5    & 0.02  0.06  6    & 0.03  0.06   3    & 0.07  0.07  4  & 0.11  0.06  11  & 0.14  0.26  2  & 0.02  0.02  2      \\ 
HIP 504  & -0.05  0.04  7   & -0.01  0.02  4  & -0.01  0.05  5    & -0.01  0.05  5    & -0.03  0.06  3    & 0.03  0.06  4  & 0.05  0.05  11  & 0.05  0.18  2  & 0.18  0.04  2     \\  
HIP 3546  & -0.09  0.07  8   & -0.15  0.04  4   & -0.06  0.07  5     & 0.01  0.06  6    & -0.09  0.05  3    & 0.05  0.06  4  & 0.10  0.06  11  & 0.17  0.35  2  & 0.01  0.04  2     \\ 
HIP 3719  & 0.01  0.06  8   & 0.05  0.03  4    & 0.01  0.03  5      & -0.01  0.04  6   & -0.10  0.03  3    & -0.02  0.07 4  & 0.03  0.05  11  & 0.03  0.10  2  & 0.29  0.07  2     \\ 
HIP 4486  & 0.02  0.07  8   & 0.07  0.01  4    & 0.04  0.07  5      & 0.01  0.06  5    & -0.02  0.03  3    & 0.03  0.06  4  & 0.08  0.05  11  & 0.09  0.22  2  & 0.15  0.08  2     \\  
HIP 6682  & 0.00  0.07  8   & 0.06  0.04  4   & 0.03  0.03  5    & 0.00  0.05  6     & -0.06  0.04  3    & -0.02  0.05 4 & 0.03  0.07  11  & 0.02  0.09   2   & 0.19  0.06  2     \\ 
HIP 7119  & -0.01  0.06  8  & 0.06  0.07  4   & 0.03  0.04  5    & 0.00  0.04  5     & -0.06  0.02  3    & -0.03  0.05 4 & 0.01  0.05  11  & 0.00  0.06   2   & 0.31  0.06  2    \\  
HIP 7710  & 0.07  0.06  8   & 0.10  0.03  4   & 0.02  0.03  5    & 0.01  0.04  6     & -0.08  0.03  3    & 0.01  0.06  4 & 0.04  0.06  11  & 0.08  0.15   2   & 0.13  0.06  2    \\  
HIP 8926  & 0.01  0.07  8   & 0.01  0.05  4   & 0.01  0.03  5    & -0.01  0.05  6    & -0.07  0.01  3    & -0.05  0.05 4 & 0.02  0.04  11  & -0.05  0.07  2   & 0.26  0.02  2    \\  
HIP 8984  & -0.03  0.05  8  & 0.07  0.06  4   & 0.01  0.04  5    & -0.01  0.05  6    & -0.11  0.03  3    & 0.02  0.05  4 & 0.05  0.06  11  & 0.09  0.13   2   & 0.19  0.05  2    \\ 
HIP 9307  & 0.03  0.07  8   & 0.07  0.06  4   & 0.06  0.04  5    & -0.01  0.06  6    & -0.05  0.04  3    & 0.00  0.06  4 & 0.02  0.05  10  & 0.06  0.05   2   & 0.23  0.06  2     \\ 
HIP 9517  & 0.14  0.05  8   & 0.07  0.09  4   & 0.08  0.03  5    & 0.06  0.07  4     & -0.13  0.05  2    & -0.05 0.05  4 & 0.01  0.07  11  & -0.11  0.00  1   & 0.28  0.05  2     \\  
HIP 11117  & -0.03  0.07  8   & -0.06  0.03  4  & 0.02  0.05  5   & -0.01  0.05  6     & -0.03  0.05  3    & 0.05  0.05  4  & 0.13  0.06  11  & 0.18  0.28  2  & -0.01  0.03  2     \\  
HIP 13786  & 0.10  0.05  8   & -0.08  0.05  4   & 0.11  0.04  5     & 0.08  0.02  6    & -0.13  0.07  3    & 0.08  0.04  4  & 0.07  0.04  11  & 0.08  0.19  2  & 0.12  0.07  2     \\  
HIP 18865  & -0.03  0.05  8   & -0.06  0.05  4  & 0.04  0.05  5     & 0.04  0.06  5    & -0.04  0.06  3    & 0.01  0.06  4  & 0.10  0.04  11  & 0.13  0.12  2  & -0.03  0.02  2       \\
HIP 19287  & -0.01  0.05  8   & -0.08  0.03  4  & -0.01  0.06  5    & 0.01  0.06  5    & -0.09  0.03  3    & 0.05  0.06  4  & 0.08  0.04  11  & 0.13  0.23  2  & 0.06  0.03  2        \\
HIP 20540  & 0.01  0.04  7    & 0.01  0.04  4   & 0.05  0.03  5     & 0.00  0.06  5    & -0.08  0.02  3    & -0.02  0.05 4  & 0.03  0.05  11  & 0.05  0.14  2  & 0.20  0.02  2      \\  
HIP 20771  & -0.07  0.07  7   & 0.01  0.03  4   & -0.03  0.03  5    & 0.01  0.06  5    & -0.06  0.03  3    & -0.03  0.05 4  & 0.05  0.05  10  & 0.05  0.11  2  & 0.17  0.06  2      \\ 
HIP 22176  & 0.02  0.08  8    & -0.10  0.05  4  & 0.11  0.06  5     & 0.06  0.07  6    & -0.05  0.03  3    & 0.08  0.08  4  & 0.11  0.07  11  & 0.13  0.29  2  & -0.01  0.05  2     \\ 
HIP 22661  & 0.01  0.08  8    & -0.11  0.06  4  & 0.08  0.04  5     & 0.01  0.06  5    & -0.06  0.05  3    & 0.16  0.06  4  & 0.14  0.07  11  & 0.24  0.48  2  & -0.09  0.01  2      \\ 
HIP 22765  & -0.06  0.05  8   & -0.05  0.03  4  & -0.03  0.05  5    & -0.02  0.06  6   & -0.04  0.04  3    & 0.06  0.05  4  & 0.08  0.04  11  & 0.14  0.26  2  & 0.09  0.04  2      \\ 
HIP 28168  & -0.02  0.07  8   & -0.06  0.05  4  & 0.05  0.06  5     & -0.01  0.06  5   & -0.02  0.04  3    & 0.04  0.08  4  & 0.08  0.06  11  & 0.15  0.37  2  & 0.01  0.07  2      \\  
HIP 28417  & 0.04  0.06  7    & 0.11  0.06  4   & 0.08  0.02  4     & -0.03  0.05  6   & -0.09  0.05  3    & 0.03  0.04  4  & 0.04  0.05  11  & 0.14  0.10  2  & 0.16  0.04  2     \\  
HIP 28677  & 0.06  0.07  8    & 0.05  0.03  4   & 0.07  0.03  5     & 0.00  0.06  5    & -0.07  0.01  3    & 0.01  0.05  4  & 0.03  0.06  11  & 0.06  0.15  2  & 0.09  0.02  2      \\  
HIP 29949  & 0.07  0.05  8    & 0.03  0.02  4   & 0.08  0.05  5     & 0.03  0.04  6    & -0.08  0.05  3    & 0.11  0.05  4  & 0.08  0.05  11  & 0.20  0.16  2  & 0.13  0.07  2     \\  
HIP 31039  & 0.01  0.08  8    & 0.04  0.01  4   & 0.02  0.03  5     & 0.00  0.06  5    & -0.09  0.02  3    & -0.04  0.05 4  & 0.02  0.06  11  & 0.08  0.10  2  & 0.16  0.03  2      \\  
HIP 32261  & -0.01  0.06  8   & 0.01  0.03  4   & 0.01  0.05  5     & 0.03  0.06  6    & -0.02  0.01  3    & 0.05  0.05  4  & 0.12  0.05  11  & 0.09  0.19  2  & 0.11  0.07  2     \\  
HIP 32844  & -0.12  0.07  8   & -0.02  0.06  4  & -0.06  0.06  5    & -0.02  0.07  6   & -0.08  0.08  3    & -0.03  0.07 4  & 0.03  0.07  11  & 0.08  0.29  2  & 0.19  0.06  2      \\  
HIP 35146  & -0.12  0.06  7   & -0.08  0.02  4  & -0.08  0.06  5    & -0.09  0.04  4   & -0.07  0.04  3    & -0.07  0.04 4  & 0.04  0.08  12  & 0.03  0.13  2  & 0.18  0.06  2     \\  
HIP 36647  & 0.01  0.05  8   & 0.06  0.04  4    & -0.01  0.04  5     & -0.06  0.04  4   & -0.18  0.06  3    & 0.03  0.05  4  & 0.04  0.06  11  & 0.24  0.11  2  & 0.13  0.07  2     \\ 
HIP 37049  & -0.01  0.06  8  & 0.09  0.05  4   & 0.01  0.04  5      & -0.02  0.06  5   & -0.07  0.03  3    & 0.01  0.06  4  & 0.03  0.04  11  & 0.05  0.15  2  & 0.21  0.05  2      \\ 
HIP 37441  & 0.06  0.06  8  & 0.08  0.04  4    & 0.10  0.03  5      & -0.01  0.06  5   & -0.07  0.02  3    & 0.05  0.04  4  & 0.01  0.03  11  & 0.02  0.08  2  & 0.27  0.03  2      \\ 
HIP 48140  & -0.11  0.08  8 & -0.01  0.03  4   & -0.10  0.04  5     & -0.07  0.05  5   & -0.09  0.04  3    & 0.00  0.05  4  & 0.03  0.07  10  & 0.08  0.18  2  & 0.10  0.07  2      \\  
HIP 48417  & -0.04  0.05  8 & -0.05  0.02  4  & -0.01  0.06  5    & -0.06  0.04  4    & -0.05  0.04  3    & -0.01  0.05 4  & 0.06  0.05  9   & 0.14  0.30  2  & 0.09  0.06  2      \\  
HIP 50526  & 0.01  0.06  8  & 0.06  0.05  4   & -0.01  0.06  5    & -0.01  0.06  6    & -0.05  0.04  2    & -0.02  0.06 4  & 0.04  0.05  11  & 0.03  0.15  2  & 0.17  0.03  2     \\  
HIP 51047  & -0.01  0.06  8 & 0.03  0.04  4   & 0.03  0.04  5     & 0.03  0.05  6     & -0.04  0.02  3    & 0.04  0.05  4  & 0.08  0.04  11  & 0.08  0.17  2  & 0.18  0.03  2      \\  
HIP 52882  & 0.03  0.07  8  & 0.06  0.04  4   & 0.04  0.04  5     & -0.01  0.04  6    & -0.08  0.05  3    & 0.03  0.05  4  & 0.03  0.04  11  & 0.09  0.26  2  & 0.20  0.05  2      \\  
HIP 58654  & 0.03  0.06  8  & 0.00  0.05  4   & 0.08  0.04  5     & 0.03  0.05  6     & -0.07  0.04  3    & 0.01  0.05  4  & 0.03  0.05  11  & 0.07  0.17  2  & 0.14  0.02  2      \\  
HIP 87629  & -0.03  0.06  6   & 0.04  0.04  4  & 0.03  0.03  5     & 0.03  0.06  6    & -0.04  0.04  3   & -0.05  0.05   4  & 0.02  0.04  11 & 0.02  0.15  2  & 0.22  0.02  2   \\   
HIP 94576  & 0.04  0.10  5    & -0.01  0.02  4 & 0.08  0.05  5     & 0.08  0.05  6    & 0.04   0.04  3   & 0.04  0.05    4  & 0.08  0.05  11 & 0.13  0.25  2  & 0.13  0.01  2     \\ 
HIP 95375  & -0.06  0.07  6   & -0.02  0.05  4 & -0.01  0.02  5    & 0.01  0.04  6    & -0.01  0.02  3   & 0.00  0.05    4  & 0.07  0.06  11 & 0.17  0.27  2  & 0.11  0.02  2      \\ 
HIP 96028  & -0.03  0.06  6   & 0.02  0.04  4  & -0.02  0.03  5    & -0.03  0.05  6   & -0.12  0.07  3   & 0.01  0.06    4  & 0.04  0.05  11 & 0.07  0.19  2  & 0.11  0.04  2    \\   
HIP 96294  & -0.02  0.07  6   & 0.06  0.05  4  & 0.04  0.03  5     & 0.02  0.05  6    & -0.06  0.06  3   & 0.03  0.07    4  & 0.06  0.04  11 & 0.08  0.20  2  & 0.17  0.04  2   \\            
HIP 102010  & 0.01  0.01  5  & 0.09  0.03  4  & 0.05  0.06  5     & -0.02  0.05  6   & -0.11  0.05  3    & 0.07  0.03   4 & 0.06  0.05  10  & 0.18  0.29   2   & 0.05  0.04  2 \\
HIP 104035  & -0.01  0.07  6 & 0.06  0.04  4  & 0.01  0.03  5     & 0.01  0.04  6    & -0.10  0.03  3    & -0.06  0.06  4 & -0.01  0.05 11  & -0.08  0.08  2   & 0.31  0.03  2  \\
HIP 105502  & -0.08  0.07  5  & 0.06  0.01  4 & -0.06  0.05  5    & -0.04  0.05  6   & -0.15  0.04  3    & -0.05  0.05  4 & 0.02  0.07  11  & 0.04  0.16  2 & 0.13  0.08  2   \\
HIP 106551  & -0.08  0.07  5  & -0.01  0.05  4 & 0.03  0.05  5     & -0.05  0.07  6   & -0.10  0.04  3    & -0.02  0.05  4 & 0.03  0.06  11  & 0.14  0.30  2 & 0.08  0.02  2  \\
HIP 107502  & 0.01  0.08  7   & -0.01  0.02  4  & 0.02  0.06  5   & -0.01  0.03  6     & -0.02  0.05  3    & 0.01  0.06  4  & 0.06  0.06  11  & 0.09  0.21  2  & 0.18  0.01  2  \\
HIP 108012  & 0.01  0.07  7   & -0.08  0.06  4  & -0.01  0.02  5  & -0.01  0.04  6     & -0.02  0.05  3    & 0.03  0.07  4  & 0.06  0.07  11  & 0.11  0.31  2  & 0.20  0.10  2    \\
HIP 108914  & -0.06  0.09  5  & -0.02  0.05  4  & 0.01  0.06  5     & 0.01  0.05  6    & -0.04  0.02  3    & 0.05  0.06   4 & 0.10  0.05  11  & 0.11  0.25  2 & 0.07  0.06  2  \\
HIP 109387  & 0.00  0.06  6   & 0.02  0.03  4   & 0.01  0.06  5     & 0.00  0.06  6    & -0.07  0.02  3    & -0.05  0.04  4 & 0.03  0.06  11  & 0.06  0.12  2 & 0.15  0.07  2   \\
HIP 109585  & -0.01  0.08  6  & 0.01  0.03  4   & 0.06  0.04  5     & 0.00  0.04  6    & -0.02  0.05  3    & 0.06  0.08   4 & 0.07  0.04  11  & 0.17  0.29  2 & 0.13  0.06  2   \\
HIP 111728  & 0.06  0.06  7   & -0.10  0.05  4  & 0.09  0.06  5   & 0.04  0.07  6      & -0.01  0.06  3    & 0.11  0.07  4  & 0.16  0.06  11  & 0.19  0.37  2  & -0.01  0.01  2   \\
HIP 113144  & 0.04  0.07  6   & 0.11  0.02  4  & 0.05  0.04  5     & -0.02  0.06  6   & -0.05  0.05  3    & 0.02  0.08   4 & 0.04  0.03  11  & 0.04  0.04  2 & 0.22  0.17  2  \\
HIP 114742  & -0.03  0.05  8  & -0.03  0.02  4  & 0.01  0.05  5   & 0.02  0.06  6      & -0.11  0.09  3    & 0.11  0.08  4  & 0.10  0.07  11  & 0.18  0.40  2  & 0.08  0.05  2    \\
HIP 115899  & -0.02  0.07  6  & 0.09  0.05  4  & -0.02  0.02  5    & -0.03  0.05  6   & -0.06  0.05  3    & -0.02  0.05  4 & 0.01  0.05  11  & 0.02  0.08  2 & 0.33  0.07  2   \\
HIP 116348  & 0.00  0.06  8   & 0.02  0.01  3   & -0.01  0.05  5  & -0.01  0.06  6     & -0.07  0.02  3    & 0.02  0.05  4  & 0.04  0.05  11  & 0.00  0.11  2  & 0.20  0.04  2    \\
{}HIP 116644  & -0.07  0.05  6  & 0.04  0.03  4  & -0.04  0.04  5    & -0.03  0.05  6   & -0.13  0.05  3    & -0.04  0.04  4 & 0.01  0.05  11  & 0.05  0.13  2 & 0.23  0.04  2   \\

\end{longtable}
\end{landscape}

\clearpage

\begin{table}
 \centering
 \begin{minipage}{140mm}
  \caption{Abundance sensitivities to various parameters for the star HIP 8926}
  \begin{tabular}{@{}lccccccc@{}}
  \hline \\
Quantity &  N &  $\sigma$(T$_{\rm eff}$)  & $\sigma$(log $g$)&$\sigma$($\xi_{\rm t}$)  &  $\sigma$([M/H])  
           &$\overline{\sigma}$(W$_{\lambda}$) &  $\sigma_{\rm model}$  \\
\\  \hline  \hline \\
{}[Na I/Fe]  &  2 &  $\pm$0.04 &  $\pm$0.01 &  $\pm$0.06 &  $\pm$0.01 &  $\pm$0.02 &  $\pm$0.08 \\
{}[Mg I/Fe]  &  4 &  $\pm$0.03 &  $\pm$0.01 &  $\pm$0.05 &  $\pm$0.00 &  $\pm$0.02 &  $\pm$0.06 \\
{}[Al I/Fe]  &  4 &  $\pm$0.04 &  $\pm$0.01 &  $\pm$0.03 &  $\pm$0.01 &  $\pm$0.02 &  $\pm$0.06 \\
{}[Si I/Fe]  &  5 &  $\pm$0.03 &  $\pm$0.06 &  $\pm$0.05 &  $\pm$0.02 &  $\pm$0.02 &  $\pm$0.09 \\
{}[Ca I/Fe]  &  8 &  $\pm$0.05 &  $\pm$0.02 &  $\pm$0.10 &  $\pm$0.01 &  $\pm$0.01 &  $\pm$0.11 \\
{}[Sc II/Fe] &  6 &  $\pm$0.01 &  $\pm$0.10 &  $\pm$0.03 &  $\pm$0.04 &  $\pm$0.02 & $\pm$0.11 \\
{}[Ti I/Fe]  &  8 &  $\pm$0.07 &  $\pm$0.01 &  $\pm$0.05 &  $\pm$0.01 &  $\pm$0.01 &  $\pm$0.09 \\
{}[Ti II/Fe] &  4 &  $\pm$0.02 &  $\pm$0.11 &  $\pm$0.12 &  $\pm$0.04 &  $\pm$0.03 &  $\pm$0.17 \\
{}[V I/Fe]   &  5 &  $\pm$0.08 &  $\pm$0.01 &  $\pm$0.04 &  $\pm$0.01 &  $\pm$0.02 &  $\pm$0.10 \\
{}[Cr I/Fe]  &  6 &  $\pm$0.05 &  $\pm$0.00 &  $\pm$0.08 &  $\pm$0.02 &  $\pm$0.02 &  $\pm$0.10 \\
{}[Mn I/Fe]  &  3 &  $\pm$0.05 &  $\pm$0.01 &  $\pm$0.05 &  $\pm$0.01 &  $\pm$0.02 &  $\pm$0.07 \\
{}[Fe I/H]   & 57 &  $\pm$0.03 &  $\pm$0.03 &  $\pm$0.08 &  $\pm$0.02 &  $\pm$0.01 &  $\pm$0.10 \\
{}[Fe II/H]  &  8 &  $\pm$0.06 &  $\pm$0.13 &  $\pm$0.09 &  $\pm$0.04 &  $\pm$0.02 &  $\pm$0.17 \\
{}[Co I/Fe]  & 4  & $\pm$0.03  & $\pm$0.03  & $\pm$0.04  &  $\pm$0.02 & $\pm$0.02  &  $\pm$0.06 \\
{}[Ni I/Fe]  & 11 &  $\pm$0.02 &  $\pm$0.04 &  $\pm$0.09 &  $\pm$0.02 &  $\pm$0.02 &  $\pm$0.10 \\
{}[Zn I/Fe]  &  2 &  $\pm$0.04 &  $\pm$0.07 &  $\pm$0.10 &  $\pm$0.03 &  $\pm$0.04 &  $\pm$0.14 \\
{}[Ba II/Fe] & 2  & $\pm$0.01  & $\pm$0.06  &  $\pm$0.13 &  $\pm$0.04 &  $\pm$0.02 &  $\pm$0.15 \\ \\
\hline  \hline
\end{tabular}
\end{minipage}
\end{table}

\begin{table*}
\centering
\begin{minipage}{180mm}
\caption{Comparisons with Pakhomov et al. (2011) for common stars. $\Delta$ represents the difference (current value - Pakhomov et al.). }
\begin{tabular}{@{}lrrrrrrrrrrrr@{}}
\hline
& HIP & HIP & HIP & HIP & HIP & HIP & HIP & HIP & HIP & HIP & HIP & \\
Quantity & 105502 & 106551 & 107502 & 108012 & 109585 & 116348 & 32844 & 35146 & 9307 & 94576 & 96028 & Mean$\pm\sigma$ 
\\ \\ \hline \hline  \\
{}$\Delta$[Na I/H] & 0.13 & 0.04 & 0.19 & $-$0.04 & 0.00 & 0.08 & $-$0.11 & 0.08 & 0.06 & 0.19 & 0.07 & 0.06 $\pm$ 0.09 \\
{}$\Delta$[Mg I/H] & 0.10 & 0.00 & 0.07 & 0.05 & 0.07 & 0.01 & $-$0.21 & $-$0.04 & $-$0.01 & 0.02 & 0.05 & 0.01 $\pm$ 0.08 \\
{}$\Delta$[Al I/H] & 0.04 & 0.06 & 0.21 & $-$0.07 & $-$0.01 & 0.11 & $-$0.1 & 0.07 & $-$0.04 & 0.06 & $-$0.05 & 0.03 $\pm$ 0.09 \\
{}$\Delta$[Si I/H] & 0.04 & 0.03 & $-$0.02 & 0.07 & $-$0.11 & 0.00 & $-$0.08 & $-$0.16 & $-$0.07 & $-$0.19 & 0.08 & $-$0.04 $\pm$ 0.09 \\
{}$\Delta$[Ca I/H] & $-$0.03 & 0.03 & $-$0.07 & $-$0.32 & $-$0.14 & $-$0.23 & $-$0.34 & $-$0.17 & $-$0.30 & 0.04 & $-$0.08 & $-$0.15 $\pm$ 0.14 \\
{}$\Delta$[Sc II/H] & 0.19 & 0.01 & 0.37 & $-$0.06 & $-$0.10 & $-$0.03 & $-$0.10 & $-$0.01 & 0.01 & 0.08 & $-$0.07 & 0.03 $\pm$ 0.14 \\
{}$\Delta$[Ti I/H] & 0.14 & $-$0.02 & 0.37 & 0.01 & $-$0.01 & 0.10 & $-$0.18 & $-$0.01 & 0.10 & 0.10 & 0.01 & 0.06 $\pm$ 0.14 \\
{}$\Delta$[V I/H] & 0.10 & $-$0.11 & 0.26 & $-$0.17 & $-$0.02 & 0.04 & $-$0.16 & $-$0.04 & 0.12 & 0.17 & $-$0.17 & 0.01 $\pm$ 0.15 \\
{}$\Delta$[Cr I/H] & 0.10 & $-$0.04 & 0.20 & $-$0.06 & $-$0.09 & 0.01 & $-$0.07 & $-$0.06 & 0.00 & 0.06 & $-$0.03 & 0.01 $\pm$ 0.09 \\
{}$\Delta$[Mn I/H] & 0.24 & 0.16 & 0.34 & 0.14 & 0.15 & 0.00 & $-$0.01 & 0.22 & 0.11 & 0.36 & 0.09 & 0.16 $\pm$ 0.12 \\
{}$\Delta$[Fe I/H] & 0.14 & 0.02 & 0.19 & $-$0.01 & $-$0.07 & 0.01 & $-$0.08 & 0.03 & 0.01 & 0.05 & 0.04 & 0.03 $\pm$ 0.08 \\
{}$\Delta$[Fe II/H] & 0.17 & 0.04 & 0.21 & $-$0.01 & $-$0.07 & 0.06 & $-$0.07 & 0.05 & 0.05 & 0.06 & 0.03 & 0.05 $\pm$ 0.09 \\
{}$\Delta$[Co I/H] & 0.07 & 0.02 & 0.26 & $-$0.02 & 0.07 & 0.08 & $-$0.06 & 0.01 & 0.05 & 0.18 & 0.04 & 0.06 $\pm$ 0.09 \\
{}$\Delta$[Ni I/H] & 0.19 & 0.07 & 0.26 & 0.04 & $-$0.03 & 0.06 & $-$0.09 & 0.07 & 0.05 & 0.07 & 0.06 & 0.07 $\pm$ 0.09 \\
{}$\Delta$[Ba II/H] & 0.34 & 0.06 & 0.55 & 0.11 & 0.02 & 0.50 & 0.26 & 0.23 & 0.26 & 0.24 & 0.14 & 0.25 $\pm$ 0.17 \\ \\ 
%\hdashline \\
$\Delta$T$_{eff}$(K) & 160 & 50 & 290 & 20 & 80 & 140 & $-$23 & 115 & 168 & 250 & 20 & 115 $\pm$ 98 \\
$\Delta$log $g$ (cgs) & 0.66 & 0.13 & 1.10 & 0.18 & 0.00 & 0.32 & 0.05 & 0.34 & 0.48 & 0.50 & 0.08 & 0.35 $\pm$ 0.33 \\
$\Delta\xi_{t}$ (km s$^{-1}$) & 0.14 & 0.3 & 0.01 & 0.14 & 0.18 & 0.14 & 0.15 & 0.00 & 0.06 & 0.14 & 0.03 & 0.12 $\pm$ 0.08 \\
\hline
\end{tabular}
\end{minipage}
\end{table*}

\clearpage

\begin{table*}
\centering
\begin{minipage}{160mm}
\caption{El/Fe for three [Fe/H] intervals : [Fe/H] = -0.2 to 0.0, 0.0 to +0.20 and +0.20 to +0.40 for our sample of Hercules stream giants 
and the local giants of Luck \& Heiter (2007)}
\begin{tabular}{@{}lcccccc@{}}
\hline \\
 
Entity&\multicolumn{2}{c} {[Fe/H] interval = -0.2 to 0.0}  &  \multicolumn{2}{c} {[Fe/H] interval = 0.0 to +0.20} 
                               & \multicolumn{2}{c} {[Fe/H] interval = +0.20 to +0.40} \\
      & Luck \& Heiter 2007 &  Hercules stream  & Luck \& Heiter 2007 & Hercules stream  & Luck \& Heiter 2007 & Hercules stream\\
      &Mean $\sigma$ N  & Mean $\sigma$ N &      Mean $\sigma$ N  & Mean $\sigma$ N     & Mean $\sigma$ N  & Mean $\sigma$ N   \\ \\ \hline \\
{}[Na/Fe] &  0.10   0.07   114 &  0.07  0.05   8 &  0.13   0.07   131 &  0.09   0.06   30 & 0.23   0.07   14 &   0.24  0.11   18  \\ 
{}[Mg/Fe] &  0.09   0.13    95 &  0.06  0.04   8 &  0.09   0.10   115 &  0.03   0.06   30 & 0.14   0.10   13 &   0.00  0.06   18  \\ 
{}[Al/Fe] &  0.10   0.08   113 &  0.16  0.05   8 &  0.08   0.05   129 &  0.11   0.07   30 & 0.13   0.07   14 &   0.13  0.06   18  \\ 
{}[Si/Fe] &  0.12   0.05   116 &  0.08  0.03   8 &  0.13   0.06   133 &  0.07   0.03   30 & 0.18   0.06   14 &   0.06  0.04   18  \\ 
{}[Ca/Fe] &  -0.03  0.07   116 &  0.07  0.03   8 &  -0.07  0.07   133 &  0.02   0.04   30 & -0.17  0.06   14 &   -0.03 0.05   18  \\ 
{}[Sc/Fe] &  -0.07  0.07   116 &  0.02  0.06   8 &  -0.13  0.08   133 &  0.02   0.03   30 & -0.16  0.10   14 &   0.12  0.43   18  \\ 
{}[Ti/Fe] &  -0.00  0.03   113 &  0.06  0.03   8 &  -0.01  0.03   132 &  0.01   0.04   30 & -0.02  0.04   14 &   -0.03 0.04   18  \\ 
{}[V/Fe]  &  -0.08  0.08   113 &  0.06  0.03   8 &  -0.04  0.08   128 &  0.01   0.04   30 & 0.06   0.10   14 &   0.01  0.06   18  \\ 
{}[Cr/Fe] &  0.02   0.09   114 &  0.01  0.03   8 &  0.03   0.04   133 &  -0.01  0.03   30 & 0.06   0.04   14 &   0.01  0.04   18  \\ 
{}[Mn/Fe] &  0.07   0.07   114 &  -0.09 0.02   8 &  0.17   0.12   131 &  -0.07  0.04   30 & 0.30   0.12   14 &   -0.05 0.04   18  \\ 
{}[Co/Fe] &  0.07   0.08   113 &  0.01  0.03   8 &  0.10   0.09   130 &  0.01   0.04   30 & 0.19   0.09   14 &   0.03  0.06   18  \\ 
{}[Ni/Fe] &  0.00   0.04   116 &  0.03  0.01   8 &  0.03   0.04   130 &  0.04   0.02   30 & 0.08   0.03   14 &   0.08  0.04   18  \\ 
{}[Zn/Fe] &  -0.04  0.29     3 &  0.05  0.07   8 &  0.04   0.07     3 &  0.08   0.07   30 & --       --    -- &   0.11  0.07   18  \\ 
{}[Ba/Fe] &  0.03   0.18   114 &  0.22  0.07   8 &  -0.10  0.18   133 &  0.24   0.08   30 & -0.37  0.17   14 &   0.14  0.10   18 \\ \\ \hline 
\end{tabular}
\end{minipage}
\end{table*}

\clearpage

\begin{figure}
\centering
\includegraphics[width=7cm, height=8cm]{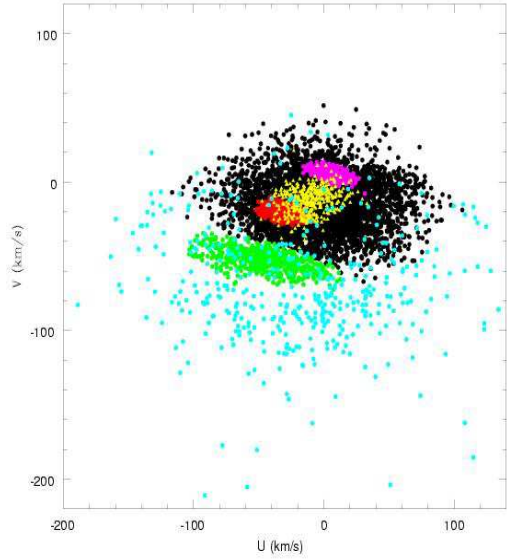}
\caption{{\it Hipparcos} giants plotted in the $U,V$-plane. Six different groups are represented. The Hercules
stream is represented by the green circles.  The Hyades-Pleiades supercluster  and the Sirius moving group are
represented by red and magenta circles, respectively. Three groups of field stars are shown: young giants by yellow circles,
high-velocity giants by blue circles and a smooth background of low-velocity stars by black circles. See Famaey et al. (2005) for detailed
descriptions of these six groups. Credit: B. Famaey et al., A\&A, 430, 165, 2005, reproduced with permission $\textcopyright$ ESO }
\end{figure}

\begin{figure*} 
\centering
\includegraphics[width=15cm, height=13cm]{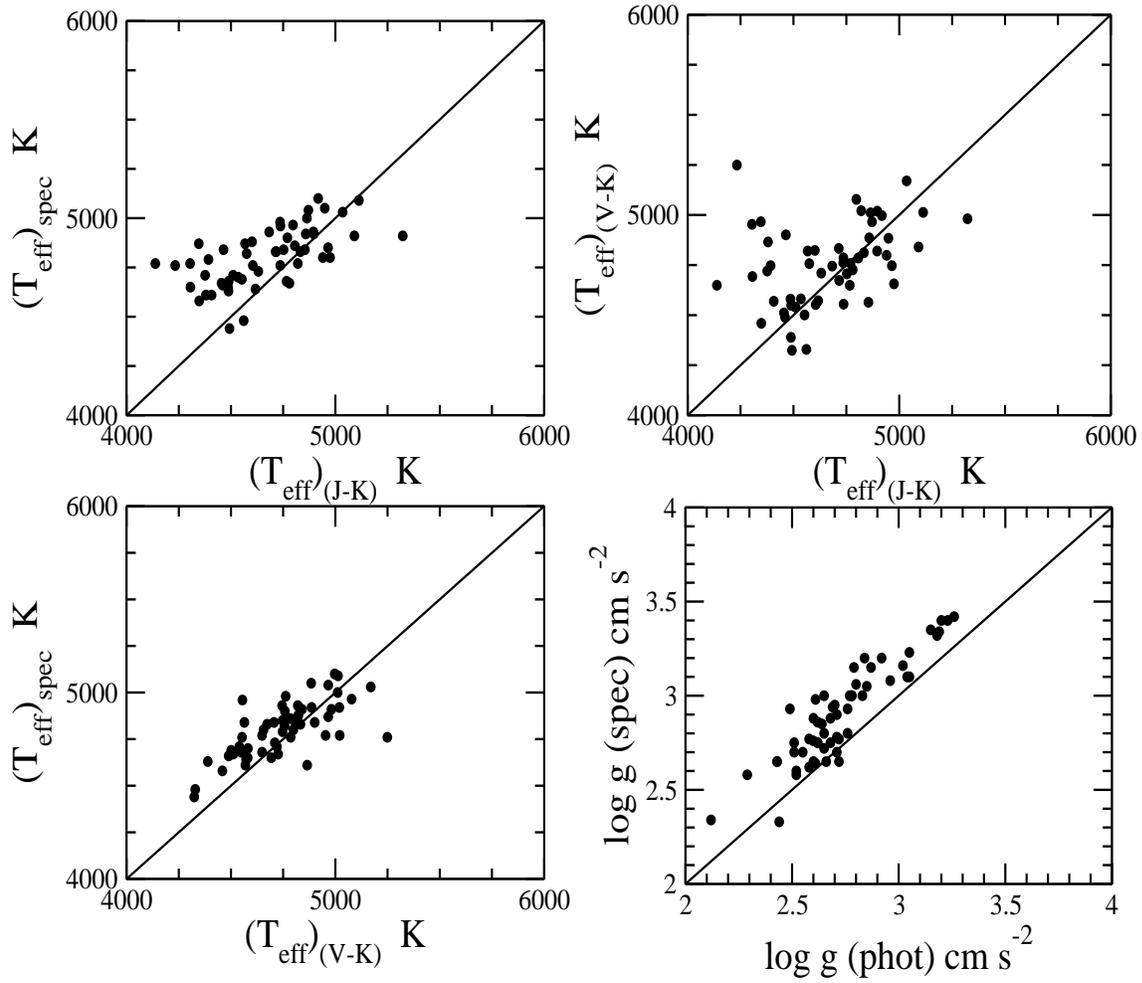}
\caption{Comparison of photometric and spectroscopic atmospheric parameters. The top two panels compare the effective temperatures
from spectroscopy ($T_{\rm eff}$)$_{\rm spec}$ with the photometric temperature from either $(J-K)$ or $(V-K)$.  The third panel 
from the top compares the
photometric temperatures from $(V-K)$ and $(J-K)$. The bottom panel compares the spectroscopic $\log g$ with that derived from
photometry. In all panels, the solid line has unit slope, i.e., x=y.}
\end{figure*}

\begin{figure*}
\begin{center}
\includegraphics[width=13cm,height=8cm]{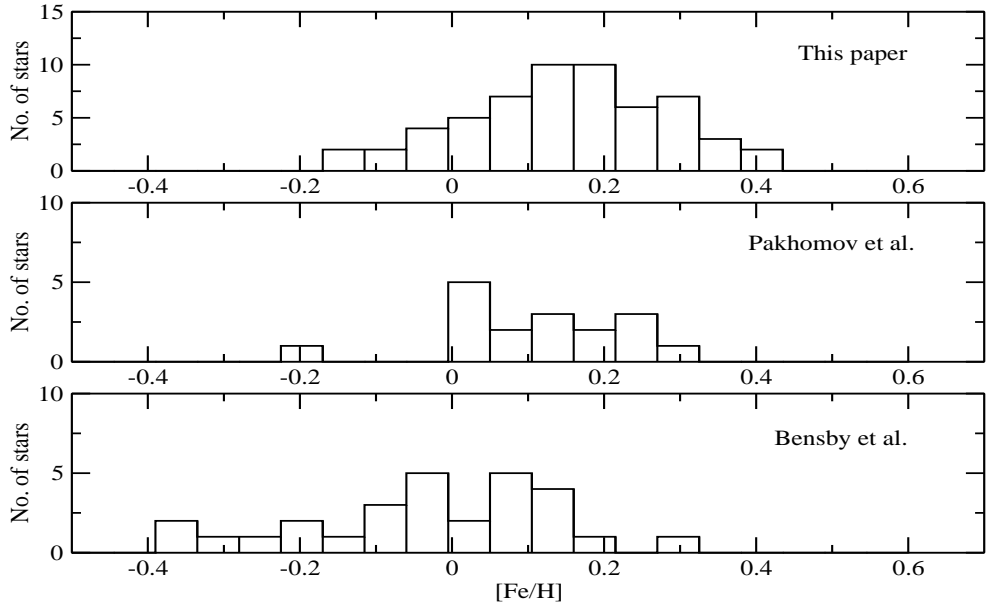}
\caption{Histograms of the Iron abundance [Fe/H] for samples of stars attributed to the Hercules stream. Top panel
shows [Fe/H] for our sample of 58 giants. The middle panel shows [Fe/H] from the 17 giants analysed by Pakhomov et al. (2011). The
bottom panel shows  [Fe/H] from the sample of dwarfs analysed by Bensby et al. (2014)  and attributed in Section 4 to the Hercules
stream.}
\end{center}
\end{figure*}.

\clearpage

\begin{figure*}
\begin{center}
\includegraphics[width=13cm, height=8cm]{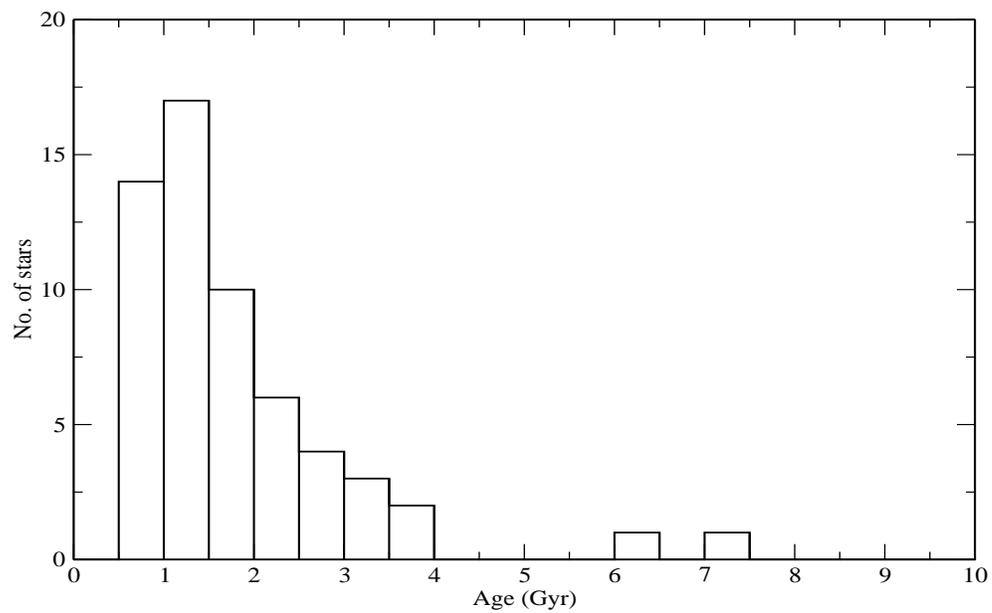}
\caption{The age distribution of the 58 giant members of the Hercules stream. Bin size = 0.5 Gyr}
\end{center}
\end{figure*}

\clearpage

\begin{figure*}
\begin{center}
\includegraphics[width=15cm, height=11cm]{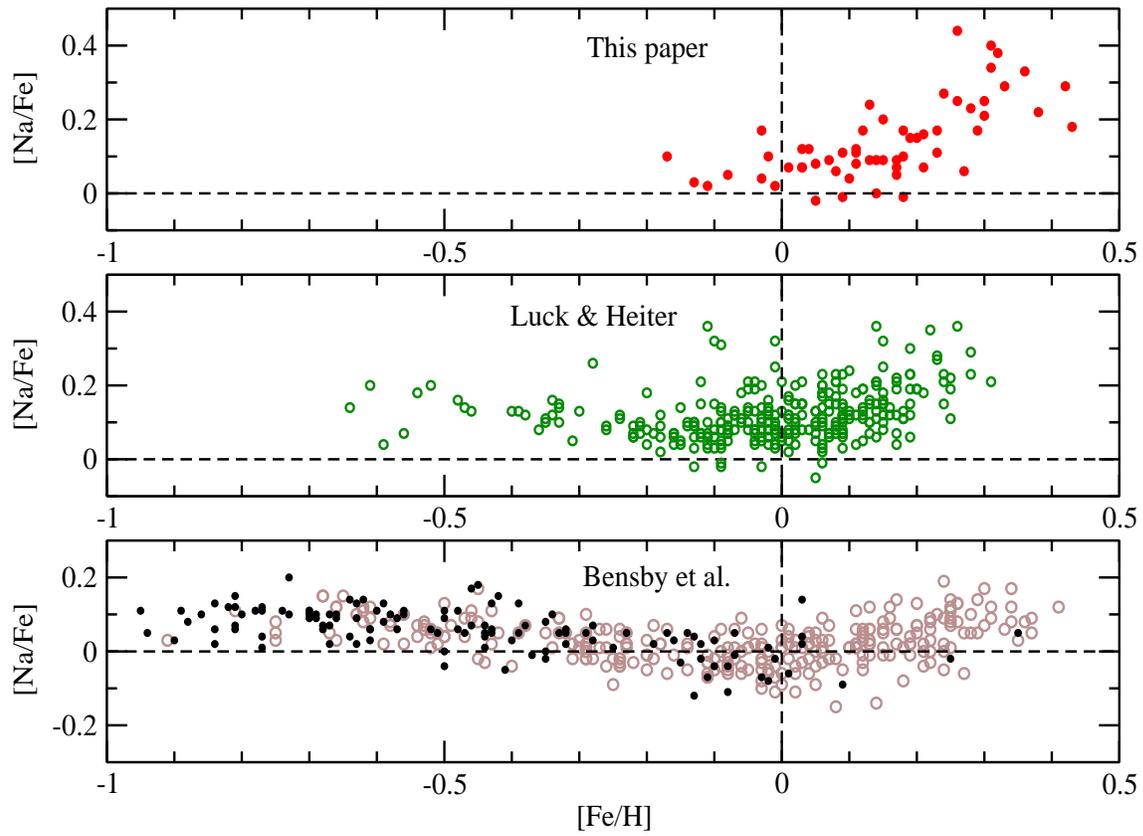}
\caption{[Na/Fe] versus [Fe/H] for our Hercules giants (top panel), local
giants analysed by Luck \& Heiter (2007) (middle panel) and thin disc dwarfs(brown unfilled circles) and thick disc dwarfs 
(filled black circles) from Bensby et al. (2014).}
\end{center}
\end{figure*}

\begin{figure*}
\begin{center}
\includegraphics[width=15cm, height=11cm]{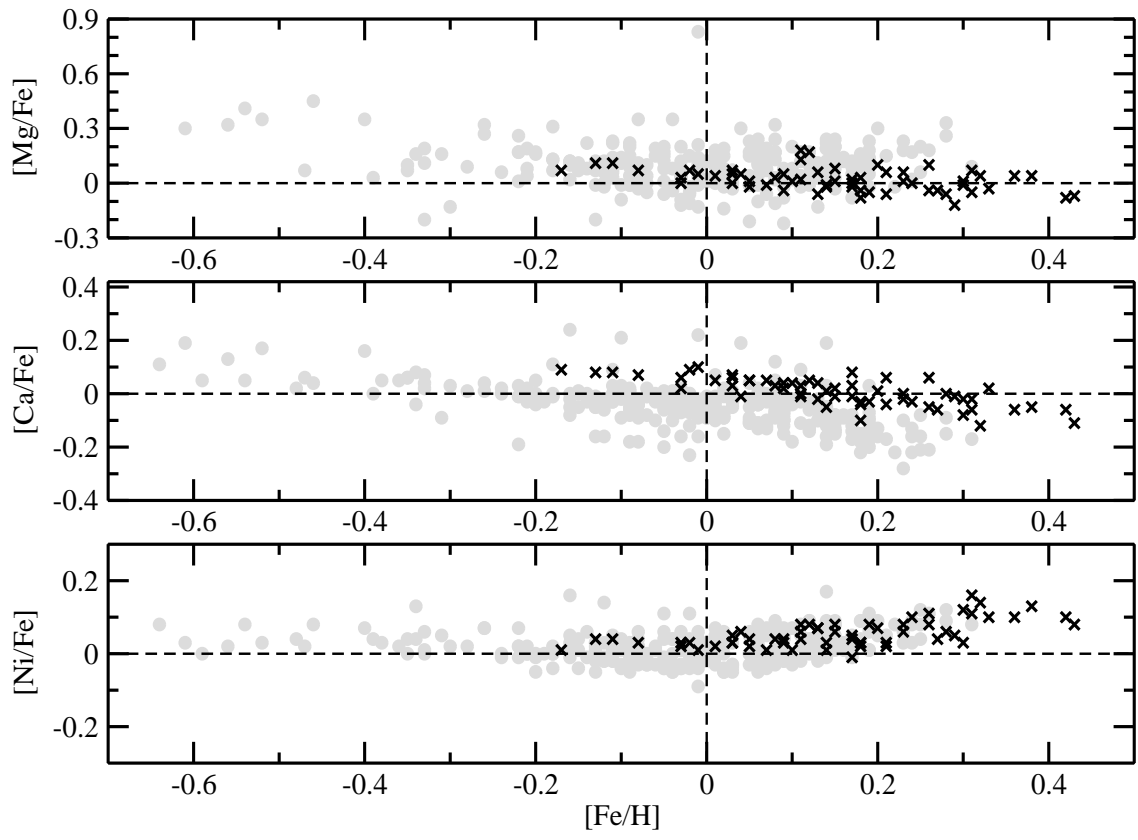}
\caption{[El/Fe] versus [Fe/H] for El= Mg, Si and Ni. In each panel, data for two samples are plotted: the Hercules stream 
giants (black crosses) and local giants (grey filled circles)from Luck \& Heiter (2007).}
\end{center}
\end{figure*}

\begin{figure*}
\begin{center}
\includegraphics[width=14cm, height=10cm]{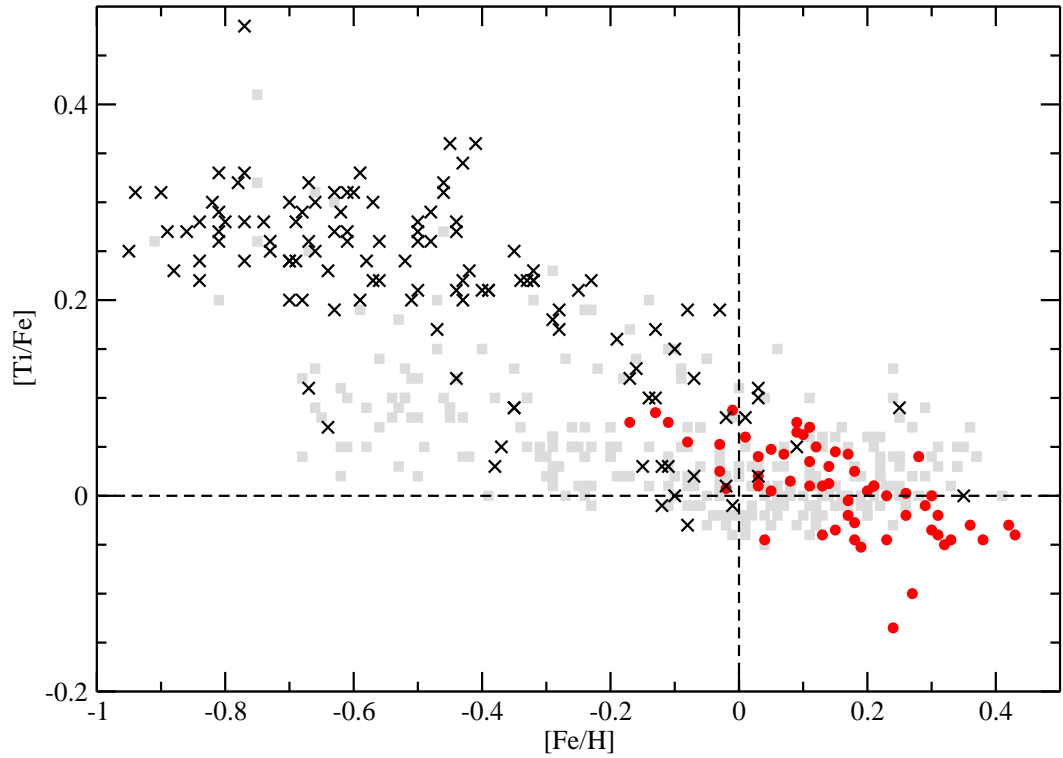}
\caption{[Ti/Fe] versus [Fe/H] for thin disc, thick disc and Hercules stream stars.  Hercules stream giants are
represented by red filled circles. Thin and thick disc dwarfs are taken from Bensby et al. (2014). Thin disc 
dwarfs with a ratio of thin to thick disc membership probability greater than 10 are represented by grey squares 
and thick disc dwarfs with a ratio of thick to thin disc membership probability of greater than 10 are represented by the black crosses.  }
\end{center}
\end{figure*}

\end{document}